\documentclass[final,1p,times,authoryear]{elsarticle}

\usepackage{float}
\usepackage{caption}

\usepackage[labelfont=bf]{caption}

\usepackage{svg}
\usepackage{tabularx} 
\usepackage{listings}
\usepackage{xcolor}

\definecolor{violetish}{RGB}{96, 0, 144}

\usepackage{amssymb}
\usepackage{lipsum}

\usepackage[round,authoryear]{natbib} 
\usepackage[colorlinks=true, linkcolor=black, filecolor=black, urlcolor=blue, citecolor=blue]{hyperref}

\usepackage{etoolbox}

\journal{Machine Learning with Applications}

\begin{document}

\begin{frontmatter}


\title{Unified Modeling Language Code Generation from Diagram Images Using Multimodal Large Language Models}

\author[ou]{Averi Bates}
\ead{averi.j.bates-1@ou.edu}

\author[ou]{Ryan Vavricka}
\ead{ryan_vavricka@ou.edu}

\author[ml]{Shane Carleton}
\ead{shane.carleton@maplarge.com}

\author[fsu]{Ruosi Shao}
\ead{rs24bl@fsu.edu}

\author[ou]{Chongle Pan\corref{cor1}}
\ead{cpan@ou.edu}

\cortext[cor1]{Corresponding author.}

\affiliation[ou]{organization={School of Computer Science, University of Oklahoma},
  addressline={110 W. Boyd St.},
  city={Norman},
  state={OK},
  country={US}}

\affiliation[ml]{organization={Enterprise Architecture, MapLarge},
  addressline={1201 Peachtree Street NE, Building 400, Suite 1750},
  city={Atlanta},
  state={GA},
  country={US}}

\affiliation[fsu]{organization={School of Communication, Florida State University},
  addressline={4100 University Center, Building C},
  city={Tallahassee},
  state={FL},
  country={US}}

\begin{abstract}
The Unified Modeling Language is a standardized visual language widely used for modeling and documenting the design of software systems. Although many tools are available that generate UML diagrams from UML code, generating executable UML code from image-based UML diagrams remains challenging. This paper proposes a new approach to generate UML code using a large multimodal language model automatically. Synthetic UML activity and sequence diagram datasets were created to train and test the model. We compared the standard fine-tuning with LoRA techniques to optimize base models. The experiments measured the code generation accuracy across different model sizes and training strategies. These results demonstrated that domain-adapted MM-LLMs perform for UML code generation automation, whereby, at the best model, it achieved BLEU and SSIM of 0.779 and 0.942 on sequence diagrams. This will enable the modernization of legacy systems and decrease the manual effort put into software development workflows.
\end{abstract}


\journal{Machine Learning with Applications}
\end{frontmatter}



\title{Unified Modeling Language Code Generation from Diagram Images Using Multimodal Large Language Models }
\let\thefootnote\relax
\footnotetext{
This is the accepted manuscript version of the article published in \textit{Machine Learning with Applications}, Volume 20, 2025, Article 100660. 
The final version is available at \url{https://doi.org/10.1016/j.mlwa.2025.100660}.
}








\section{Introduction}
\label{introduction}

In software engineering, the Unified Modeling Language (UML) is widely used to represent and analyze complex systems visually \citep{Booch1998TheUM}. UML provides a standardized framework for documenting, designing, and refining software solutions \citep{Lu2023SystematicRO,metzner2024systematic}. It supports various stages of the development lifecycle by structuring software design and clarifying relationships among components, encouraging collaboration among stakeholders \citep{uml_review}. Metzner compares UML proficiency to an architect’s use of blueprints, highlighting its importance in creating systematic, organized design frameworks \citep{metzner2024systematic}.

Despite its utility, UML diagrams are often stored in fixed formats such as PDFs or images, limiting their usability in automated workflows. Static formats complicate iterative design tasks, such as design verification or consistency checks, particularly in agile development. Moreover, in legacy systems, UML diagrams and corresponding code may become inaccessible, leading to outdated documentation and increasing manual reconstruction efforts. For instance, a bank modernizing its legacy transaction processing system may only have outdated PDFs and scattered notes of its original UML diagrams. Without tools to automate conversion into editable, machine-readable formats, developers must painstakingly reconstruct designs, risking overlooked dependencies, security vulnerabilities, and higher maintenance costs. Similar challenges arise in collaborative environments where static UML diagrams lose relevance over time as project teams evolve. The need for automated UML-to-code generation is becoming increasingly urgent as systems grow in complexity. Manual translation of diagrams into code is labor-intensive and error-prone. Automated tools that extract and convert UML diagrams into editable formats can streamline development, reduce errors, and preserve design integrity.

\subsection{Addressing the Problem}

To address the limitations of rule-based tools, this research leverages Multimodal Large Language Models (MM-LLMs), which integrate advanced machine learning techniques to process visual and textual data. MM-LLMs analyze complex UML diagram components without relying on static rule sets, such as class relationships, dependencies, and hierarchical structures. MM-LLMs can handle non-standard, hand-drawn, or domain-specific diagrams by dynamically interpreting visual UML constructs and corresponding annotations. The multimodal approach offers significant advantages for scalability and adaptability. As UML diagrams evolve during iterative development, MM-LLMs can extract meaningful information from incomplete or partially updated diagrams, ensuring consistency across workflows like design validation, automated code generation, and documentation.

Automated code generation from visual representations has gained traction in recent years. For example, pix2code demonstrated the conversion of GUI screenshots into executable code across platforms like iOS, Android, and the web \citep{beltramelli2018pix2code}. Using CNNs and LSTM networks, it bypasses complex heuristics by interpreting GUI elements directly from images. Building on this work, \citet{Cai2023ANC}’s GUICG enhances component localization and classification through CNN-based target detection, achieving notable accuracy improvements on large-scale GUI datasets.
Beyond GUIs, Ellis et al. explored code generation from freehand sketches using a two-stage model combining CNNs with program synthesis techniques \citep{ellis2018learning}. Identifying geometric primitives and constructing structured code representations highlights the potential of machine learning in bridging visual and code-based workflows. Researchers have also started developing models that extract elements from UML diagrams stored in non-editable formats. For instance, ReSECDI identifies core UML components like classes and relationships across varying formats \citep{CHEN2022111431}. However, while ReSECDI provides syntactic extraction, it does not yet generate executable code, underscoring the need for machine learning techniques to bridge this gap \citep{conrardy2024}.

Large Language Models (LLMs) have demonstrated impressive capabilities in understanding and generating human-like text. Advanced transformer-based architectures such as BERT \citep{devlin2018bert}, GPT \citep{radford2018improving}, and T5 \citep{raffel2020exploring} capture contextual relationships within text, enabling tasks like summarization, question answering, and code generation. However, LLMs process only textual data, limiting their utility in domains requiring multimodal understanding, such as image or video analysis. Researchers developed MM-LLMs that incorporate visual, auditory, and textual data into unified frameworks to address these limitations. Models like CLIP \citep{radford2021learning} align visual embeddings with textual embeddings, enabling cross-modal tasks such as visual question answering (VQA) and zero-shot image classification. For example, CLIP (Contrastive Language–Image Pretraining) learns shared visual and textual semantics from large-scale datasets of image-text pairs, effectively bridging linguistic context and visual perception. MM-LLMs leverage these multimodal capabilities to tackle tasks that traditional LLMs cannot address, such as interpreting UML diagrams and generating corresponding executable code. MM-LLMs align diagram structures with text-based outputs by combining visual encoders with language models, enhancing accuracy and flexibility for applications like UML-to-code automation.

\subsection{Research Objectives and Contributions}

Our research aims to enhance the capabilities of MM-LLMs for UML-to-code generation by fine-tuning them to handle activity and sequence diagrams—key behavioral diagrams in software design. By leveraging MM-LLMs' ability to interpret visual UML components and generate executable code, this work addresses the limitations of existing tools. Specifically, the study develops a scalable framework to evaluate model performance across dataset sizes and UML complexities. The key contributions include a framework for UML diagram-to-code automation, insights into factors affecting MM-LLM generalization, and recommendations for future enhancements, such as improving dataset diversity and multimodal alignment. This research bridges the gap between static UML documentation and modern, automated workflows, offering practical solutions for improving software design efficiency.

\subsection{UML Scope}
UML serves as a standard framework for visualizing and analyzing system designs. It consists of various diagram types categorized into structural and behavioral groups. Structural diagrams, such as class diagrams, capture static system elements like classes, attributes, and relationships. On the other hand, behavioral diagrams focus on dynamic interactions and workflows within systems. Our research focuses on behavioral diagram types—activity and sequence diagrams—because they are critical in representing workflows and communication flows, which are essential for UML-to-code automation.

\begin{figure}[htbp]
    \centering
    \includegraphics[width=0.8\textwidth]{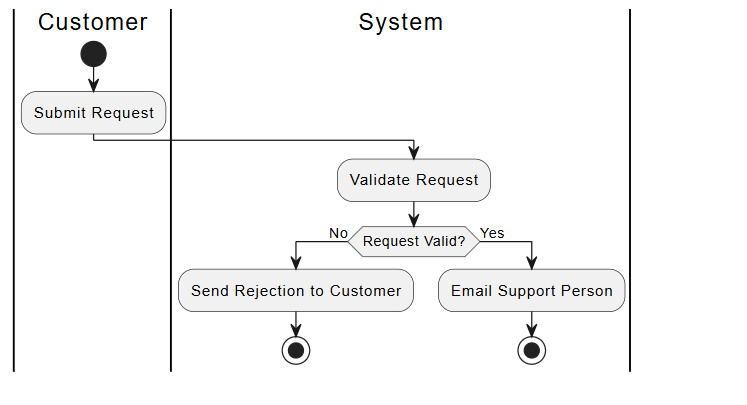}
    \caption[Customer Request Activity Diagram]{Activity diagram showing the decision paths for valid and invalid requests.}
    \label{fig:activity_example}
\end{figure}

\begin{figure}[htbp]
    \centering
    \includegraphics[width=0.7\textwidth]{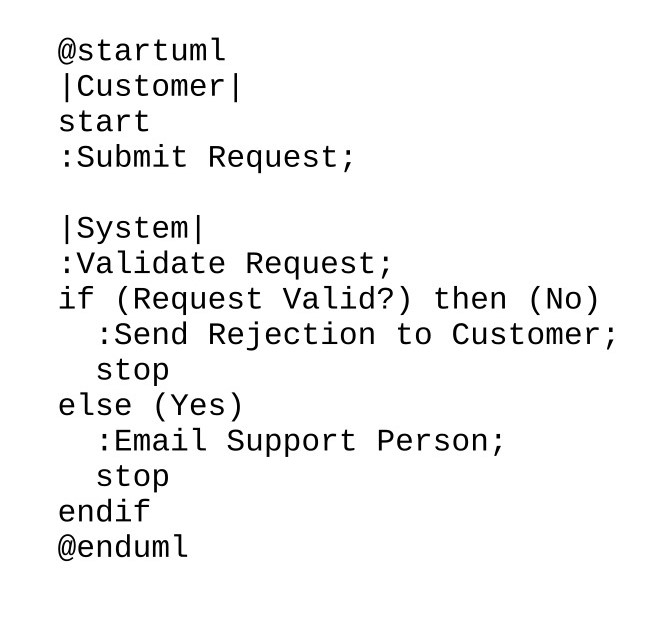}
    \caption[Customer Request Activity Code]{PlantUML code that represents the activity diagram structure, including the decision points and action paths.}
    \label{fig:activity_example_code}
\end{figure}

Activity diagrams excel at modeling control flow and decision-making processes within a system, providing a high-level overview of workflows. Each action or task is depicted as a rounded rectangle, with arrows (control flows) indicating the progression of tasks. Decision nodes handle conditional branches, while swimlanes assign responsibilities to actors or components. These features make activity diagrams particularly valuable for visualizing multi-actor systems and concurrent processes. Fig \ref{fig:activity_example} provides an example of an activity diagram for a request validation process, where customer requests are either approved or rejected based on conditions. The corresponding PlantUML code seen in Fig \ref{fig:activity_example_code} defines this workflow and its components.

\begin{figure}[htbp]
    \centering
     \includegraphics[width=0.7\textwidth]{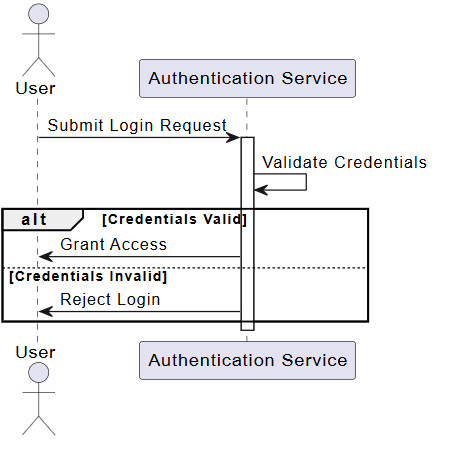}
    \caption[Website User Sequence Diagram]{Sequence diagram representing validating user information and granting access.}
    \label{fig:sequence_example}
\end{figure}

\begin{figure}[htbp]
    \centering
    \includegraphics[width=0.85\textwidth]{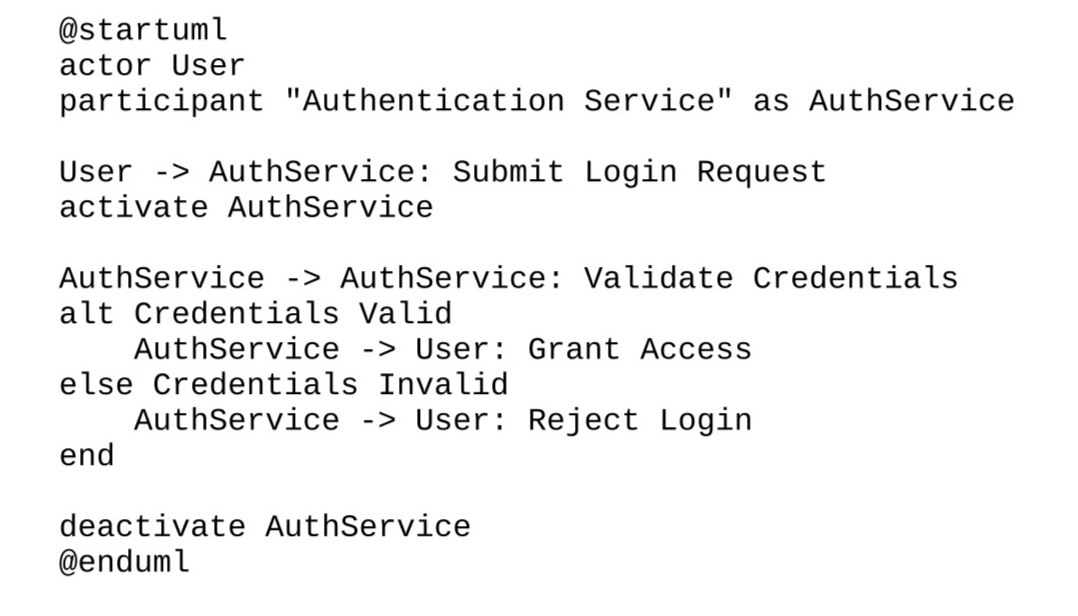}
    \caption[Website User Sequence Diagram Code]{PlantUML code that defines the interactions for the user login process shown in the sequence diagram.}
    \label{fig:sequence_example_code}
\end{figure}

In contrast, sequence diagrams focus on temporal interactions between system components, emphasizing communication order and dependencies. A lifeline represents each component, while horizontal arrows signify message exchanges or method calls. Sequence diagrams excel in modeling client-server interactions, API calls, and service workflows, offering developers insight into component dependencies over time. Figs \ref{fig:sequence_example} and \ref{fig:sequence_example_code} present a user log-in process captured in a sequence diagram. It models how users interact with the authentication service, highlighting decision paths for valid and invalid credentials. The corresponding PlantUML code details these interactions, focusing on the sequence of events and activation periods.

Activity diagrams and sequence diagrams are inherently complementary. While activity diagrams provide a broad, workflow-centric perspective suited for process automation, sequence diagrams emphasize fine-grained, event-driven interactions between components. Focusing on these two diagram types, allows actionable insights into UML-to-code automation. Activity diagrams support the generation of control-flow logic and parallel execution, while sequence diagrams align with detailed communication mapping.

\subsection{Automating Code Generation}
LLMs have significantly advanced automated code generation, providing tools that analyze textual and structural information to generate functional code. For example, Meta’s CodeCompose offers developers real-time code suggestions and streamlining routine coding tasks through context-aware predictions \citep{10.1145/3643774}. Similarly, OpenAI’s Codex, fine-tuned from GPT-3 on GitHub datasets, achieves a notable success rate in solving coding challenges through iterative sampling strategies \citep{chen2021evaluatinglargelanguagemodels}. These tools demonstrate the potential of LLMs to improve productivity by reducing manual intervention in code generation. Despite their success, traditional LLMs are limited in diagram-to-code automation. Converting UML diagrams into executable code demands both linguistic and visual comprehension. Thus, the use of MM-LLMs
fine-tuning techniques play a critical role in enhancing MM-LLMs for UML-to-code generation. Instruction-based fine-tuning and Chain-of-Thought prompting improve the models' ability to generate contextually accurate and logically consistent code. Recent studies have shown that fine-tuned models generalize effectively across benchmarks, maintaining structural fidelity in generated outputs \citep{chung2022scalinginstructionfinetunedlanguagemodels}. Studies like Plot2Code and Design2Code highlight the challenges of converting structured visuals into functional code. For instance, while GPT-4 achieved high accuracy in generating executable code from scientific plots, its performance declined with increasing visual complexity \citep{shi2024chartmimicevaluatinglmmscrossmodal, wu2024plot2code}. These findings underscore the importance of tuning MM-LLMs to handle intricate UML diagrams, where maintaining relational and sequential accuracy is critical. In this research, we extend these advancements by fine-tuning MM-LLMs to generate code from activity and sequence diagrams autonomously. The models are trained to align diagram structures with functional outputs, ensuring high syntactic and structural fidelity.

\section{Methodology}
Automating UML code generation requires models capable of processing visual and textual data. We use the Large Language and Vision Assistant (LLaVA) \citep{liu2024visual}, specifically its improved version, LLaVA-1.5 \citep{liu2024improved}, a foundational model. Both LLaVA and LLaVA-1.5 integrate a CLIP-based visual encoder \citep{10647787} with the Vicuna language model \citep{vicuna2023} via a vision-language connector, enabling the extraction of visual features from UML diagrams and translating them into accurate textual outputs. LLaVA uses a linear projection layer to align visual embeddings with the language model. In contrast, LLaVA-1.5 introduces a multi-layer perceptron as a more complex connector, enabling nuanced interactions between visual and textual representations. The architectural enhancements, combined with expanded visual instruction tuning on VQA and diagram datasets, improve LLaVA-1.5’s performance for tasks requiring detailed diagram analysis, such as UML-to-code generation. 

\subsection{Fine-tuning and LoRA Fine-tuning}

The base model, the 'original model' or 'baseline model,' is the pre-trained LLaVA model without fine-tuning. Although this base model has strong general capabilities, it struggles with the task-specific syntax and structure required to accurately interpret and generate UML diagrams, particularly given the detailed nature of some UML formats. Fine-tuning addresses this limitation by training the model on task-specific data, enabling it to understand better the unique demands of UML diagrams, such as their precise syntax and structural rules. Our study uses 'full fine-tuning' and 'standard fine-tuning' interchangeably to describe tuning that updates all model weights using additional, task-specific data. The baseline model, which remains unaltered, serves as the starting point for the fine-tuning process. Fine-tuning adapts the model to address UML diagram-specific requirements better, improving its performance and accuracy in diagram-to-code generation tasks.

Four models were trained using two fine-tuning strategies: standard fine-tuning and LoRA (Low-Rank Adaptation) fine-tuning. The models included two sizes: one with 7 billion parameters (7B) and one with 13 billion parameters (13B). The model sizes represent different trainable parameters, with the large model having more. Standard fine-tuning updates all model parameters, while LoRA introduces trainable low-rank matrices that reduce the computational burden by focusing on updating only selected layers. LoRA allows for efficient adaptation of large language models, such as LLaVA-1.5. 

LoRA reduces the number of trainable parameters by introducing trainable low-rank matrices into the architecture. Specifically, instead of updating all parameters in the model, LoRA modifies only a subset of the parameters in the form of low-rank matrices, significantly decreasing memory consumption during training. The low-rank approximation minimizes the number of trainable parameters by focusing only on updating specific components, leading to substantial reductions in memory usage and computational overhead \citep{hu2021loralowrankadaptationlarge}. Despite this simplification, LoRA can demonstrate performance comparable to full fine-tuning in several large-scale language models. Given the scale of models like LLaVA-1.5, incorporating LoRA enables the system to maintain computational efficiency without compromising its ability to adapt to the unique requirements of tasks such as UML diagram analysis. This balance between resource efficiency and adaptability makes LoRA a compelling approach for fine-tuning large models on domain-specific tasks.

\subsection{Hyperparameters}

The fine-tuning process utilized several hyperparameters. Common hyperparameters, such as the \textit{learning rate} and \textit{batch size}, applied to both standard and LoRA fine-tuning, while others were specific to LoRA. The differentiation ensured that each fine-tuning strategy was optimized for its specific methodology. The process adhered to the original implementation of LLaVA-1.5 to represent the architecture accurately. Table \ref{table:finetuning} summarizes the hyperparameters applied in both fine-tuning strategies.

\begin{table}[htbp]
\centering
\caption{Hyperparameters for LoRA and Standard Fine-tuning}
\label{table:finetuning}
\begin{tabularx}{\linewidth}{l X X} 
 \hline
 \textbf{Hyperparameter} & \textbf{LoRA Fine-Tuning} & \textbf{Standard Fine-Tuning} \\ \hline
 lora\_enable                & True            & None        \\ 
 lora\_r                     & 128             & None        \\ 
 lora\_alpha                 & 256             & None        \\ 
 mm\_projector\_lr           & $2 \times 10^{-5}$ & None        \\ 
 learning\_rate              & $2 \times 10^{-4}$ & $2 \times 10^{-5}$ \\ 
 per\_device\_train\_batch\_size & 16              & 8           \\ 
 gradient\_accumulation\_steps  & 4              & 4           \\ 
 \hline
\end{tabularx}
\end{table}

The hyperparameters \textit{lora\_r} and \textit{lora\_alpha} are central to LoRA fine-tuning. \textit{lora\_r} determines the rank of the low-rank matrices integrated into the model. Increasing \textit{lora\_r} enhances the expressive power of these matrices, enabling more nuanced adaptations but requiring greater memory and computational resources \citep{hu2021loralowrankadaptationlarge}. \textit{lora\_alpha}, a scaling factor, adjusts the influence of the low-rank matrices during training. Increasing \textit{lora\_alpha} amplifies the impact of the matrices on the model output, enhancing their prominence in the learning process. The \textit{mm\_projector\_lr} hyperparameter sets the learning rate for the projector layer. Since LoRA fine-tuning updates only a subset of parameter values, the projector learning rate ensures balanced updates concerning the rest of the model. LoRA fine-tuning employs a higher learning rate and larger batch size, as updating fewer parameters than standard fine-tuning allows the model to converge more efficiently. A batch size of 8 was used for standard fine-tuning, while LoRA fine-tuning utilized a batch size of 16 to accommodate its more efficient parameter updates. Despite the differences in batch size, the training times and computational performance metrics provide valuable insights into each approach's efficiency and scalability.

\subsection{Synthetic Data Generation}

Two types of synthetic UML diagrams—activity and sequence diagrams—were programmatically generated using PlantUML, a text-based diagramming tool widely adopted for its ease of integration and automation. The diagrams were constructed using randomized text strings drawn from a curated vocabulary of the 25,286 most popular words found in English. These strings were inserted into standard PlantUML syntax structures, thereby generating a wide variety of visually valid UML diagrams that are structurally diverse but semantically nonsensical. Importantly, the diagrams do not follow any coherent application logic; rather, they serve as syntactic approximations of real UML diagrams. This approach was chosen to isolate the visual parsing and language modeling capabilities of the system without introducing domain-specific priors.

\begin{figure}[htbp]
    \centering
    \includegraphics[width=0.7\textwidth]{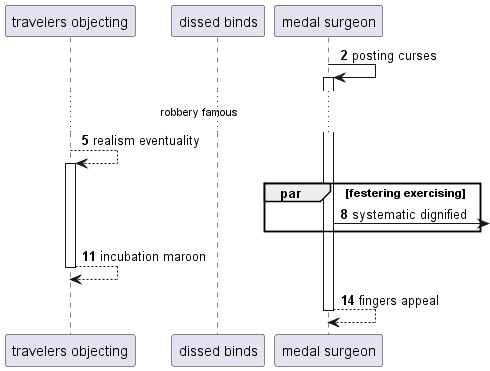}
    \caption{Sequence diagram depicting participants, messages, and activation events.}
    \label{fig:sequence_diagram_combo}
\end{figure}

\begin{figure}[htbp]
    \centering
    \includegraphics[width=0.9\textwidth]{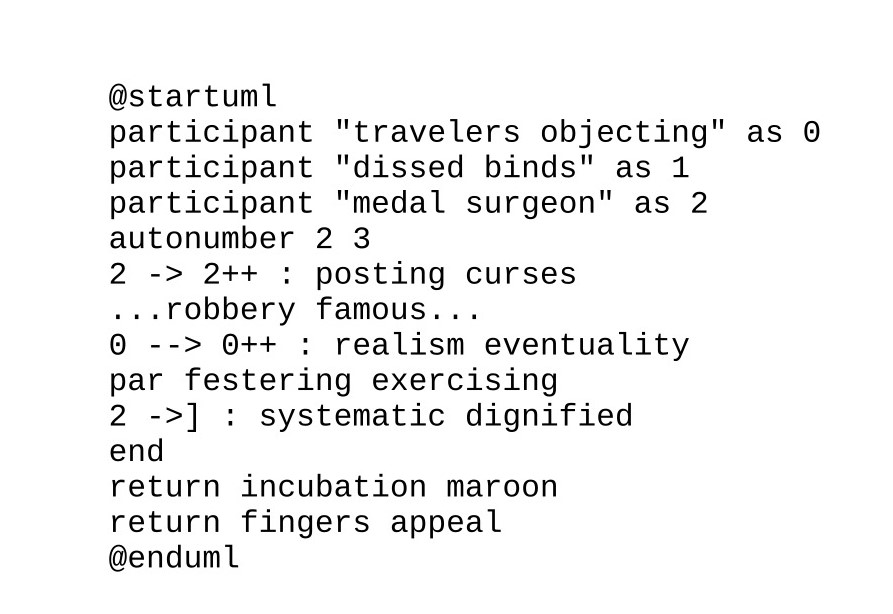}
    \caption{PlantUML code generating the sequence diagram.}
    \label{fig:sequence_code}
\end{figure}

PlantUML enables direct conversion of textual UML descriptions into graphical formats, supporting both rasterized PNG and vectorized SVG outputs. Each synthetic diagram was exported in PNG format. The sequence diagrams emphasize participant interactions and message flows, as illustrated in Fig \ref{fig:sequence_diagram_combo}. The corresponding PlantUML code showcases the structured textual representation of visual relationships in Fig \ref{fig:sequence_code}.  In a similar fashion, activity diagrams were synthesized to reflect control flows including branching, merging, and parallel operations, represented visually in \textbf{Fig. S1} and in PlantUML text format (\textbf{Fig. S2}). Their labels and process steps are composed of randomly selected words, resulting in workflows that lack meaningful interpretation. Nevertheless, they adhere to the syntactic rules of UML activity diagramming, making them suitable for testing model robustness to surface-level structural variation.

\subsubsection{Dataset Breakdown}

The synthetic data set consisted of activity and sequence diagrams split into four sizes, 'small', 'medium', 'large', and 'extra large', each containing various diagrams and the corresponding UML code files. Each one had varying amounts of training and testing data. Table\ref{table:dataset_breakdown} outlines the distribution of the data set for training and testing. The training was conducted separately on the types of activity and sequence diagram to optimize the model's performance, ensuring task-specific optimization.

\begin{table}[htbp]
\centering
\caption{Dataset Sizes for Training and Testing}
\label{table:dataset_breakdown}
\begin{tabularx}{\linewidth}{l X X X}
\hline
\textbf{Dataset Size} & \textbf{Diagram Type} & \textbf{Training Size} & \textbf{Testing Size} \\ \hline
Small                 & Activity             & 6,000                  & 1,500                \\ 
Medium                & Activity             & 12,000                 & 3,000                \\ 
Large                 & Activity             & 24,000                 & 6,000                \\ 
Extra Large           & Activity             & 120,000                & 30,000               \\ 
Small                 & Sequence             & 6,000                  & 1,500                \\ 
Medium                & Sequence             & 11,994                 & 2,998                \\ 
Large                 & Sequence             & 23,994                 & 5,998                \\ 
Extra Large           & Sequence             & 119,994                & 29,998               \\ \hline
\end{tabularx}
\end{table}

Each dataset entry contains:
\begin{enumerate}
    \item UML diagram image (PNG/SVG format).
    \item Corresponding UML code (PlantUML syntax text file).
    \item Pairing of prompt templates.
\end{enumerate}

All generated diagrams and corresponding PlantUML source files were stored with unique identifiers to ensure dataset traceability and to prevent unintentional duplication. Additionally, each image-code pair was serialized into a JSON format (see Supplemental \textbf{Fig. S3}), which was used as the primary input format during model training. The dataset was partitioned using a simple random sampling procedure to produce an 80/20 train-test split. Testing examples were selected without regard to structure or content, ensuring unbiased evaluation across the synthetically generated samples. The testing split had an additional step where data was organized in a question-answer format and added as another JSON file. The structured format supports effective multimodal learning, as demonstrated below.

\subsubsection{Real-World Testing Evaluation}
While the majority of our dataset is synthetically generated to ensure controlled variability and sufficient volume for training, it is important to assess how well models perform on naturally occurring data. To this end, we curated a small set of real-world UML diagrams—29 activity diagrams and 28 sequence diagrams—totaling 57 examples. These were sourced from publicly available resources and serve as a benchmark for evaluating the generalization capabilities of models trained exclusively on synthetic data. The real-world data is used solely for testing and enables comparison across models with varying levels of training exposure. We evaluate the original model, a minimally trained small LoRA variant, two standard models (7B and 13B) trained on the small dataset, and finally, models trained extensively on the extra-large dataset using both LoRA and full parameter tuning. This setup allows us to examine the influence of training intensity—from no training to full-scale training—on the model’s ability to generalize to naturally structured UML diagrams. Including this evaluation provides valuable insights into the robustness and practical applicability of our models. It helps determine whether large-scale synthetic training introduces limitations or enhances the model’s ability to interpret real-world inputs. The distribution of the real-world testing dataset is summarized in Table \ref{table:realworld_testing}. 

The real-world UML diagrams were sourced from publicly available, educational, and technical resources that include software engineering tutorials, documentation repositories, and open-source guides. Specifically, we collected data from three primary sources: the official PlantUML guide \citep{plantuml2025guide}, as well as tutorial-based resources on activity and sequence diagrams from WebDevTutor \citep{webdevtutor2023activity, webdevtutor2023sequence}. Selection criteria focused on clarity, diversity of structure, and representational fidelity to practical use cases in software design. All chosen diagrams were fully rendered and accompanied by the corresponding PlantUML source code, ensuring their suitability for both image-based and code-based model evaluation. Unlike the synthetic diagrams, which were generated by inserting randomized text strings into grammatically correct UML syntax templates, the real-world diagrams were authored by domain experts or educators to illustrate concrete workflows and interactions. These real diagrams typically exhibit meaningful semantic content, coherent control or message flows, and intentional labeling conventions that reflect real application logic or business processes. For instance, activity diagrams in the real-world set often include process names suggesting a clear operational sequence. Sequence diagrams show realistic inter-object communications such as login procedures or database queries, in contrast to the semantically nonsensical but syntactically valid interactions found in synthetic examples.

\begin{table}[htbp]
\centering
\caption{Real-World Testing Data Distribution}
\label{table:realworld_testing}
\begin{tabular}{l c}
\hline
\textbf{Data Type} & \textbf{Testing Size} \\ \hline
Activity Diagrams  & 29                    \\
Sequence Diagrams  & 28                   \\ \hline
\end{tabular}
\end{table}

\subsection{Hardware and Pipeline Devoplment}
We completed all fine-tuning and evaluation tasks on the University of Oklahoma’s supercomputing infrastructure, leveraging an SLURM-managed HPC cluster. Depending on node availability, training was conducted on NVIDIA A100 GPUs, either 40GB or 80GB versions. For standard full-parameter fine-tuning—required for larger models like Vicuna-13B and LLaVA-13B—we utilized four A100 GPUs to meet the memory and compute demands. In comparison, LoRA fine-tuning was successfully executed using just two A100 GPUs. All jobs were submitted via SLURM scripts, which specified GPU resources, runtime configurations, and distributed training parameters.

Our approach followed the official LLaVA training pipeline from the project’s GitHub repository, incorporating full model and LoRA-based fine-tuning scripts. The process involved two key stages: feature alignment and visual instruction tuning. Depending on the experiment, we used either the standard (fine-tune.sh) or LoRA (finetune\_lora.sh) script, with only minimal modifications tailored to our HPC setup and resource allocations  \citep{liu2024improved} \footnote{Code available at: \url{https://github.com/haotian-liu/LLaVA}}. Model checkpoints were saved at predefined intervals, and SLURM output was used to monitor logs in real-time. We performed all inference runs on a single A100 GPU per model for consistent evaluation, applying this to both our trained checkpoints and baseline models. The setup ensured uniform memory and compute conditions across all experiments. All associated pre- and post-processing—such as JSON parsing and results aggregation—was handled through Python 3.11 scripts within a Linux terminal environment.

\subsection{Evaluation Metrics}
We hypothesize that using large datasets and larger model sizes will improve BLEU and SSIM metrics performance, reflecting enhanced textual accuracy and visual fidelity in generated UML code. Simultaneously, we expect these configurations to yield lower error rates, indicating fewer syntax errors, structural misalignments, and overall diagram inaccuracies. While both LoRA fine-tuning and standard fine-tuning are predicted to produce comparable BLEU and SSIM improvements, LoRA is anticipated to offer superior computational efficiency. Specifically, it should achieve faster training times and require fewer computational resources while maintaining similar output quality. This hypothesis underpins the study's goal of balancing accuracy and efficiency to optimize UML-to-code generation workflows.

The BLEU metric, introduced by Papineni et al. \citep{bleu2002}, is widely used for evaluating text generation tasks. It measures the similarity between generated text and a reference by comparing n-grams, offering a quantitative measure of syntactic accuracy. BLEU's simplicity, scalability, and independence from specific programming languages make it well-suited for large-scale evaluation of generated UML code. In prior studies, BLEU has been successfully applied to similar tasks such as evaluating large-scale language models for code generation \citep{chen2021evaluatinglargelanguagemodels,zhang2023planninglargelanguagemodels}. A higher BLEU score indicates stronger alignment between the generated output and the ground truth, ensuring that syntactic fidelity is maintained during UML-to-code translation.

The Structural Similarity Index Measure (SSIM) is a common image quality metric that evaluates the visual similarity between generated and reference images \citep{1284395, Reznik2023AnotherLA}. Unlike pixel-based metrics, SSIM considers luminance, contrast, and structural features to align with human visual perception. This makes SSIM especially relevant for UML diagrams, as it prioritizes the layout and relationships between elements over minor pixel variations. SSIM produces scores ranging from -1 to 1, where higher values reflect closer visual resemblance. When evaluating UML diagrams, SSIM helps ensure that structural elements, such as object relationships, decision nodes, and component alignment, remain intact. \textbf{Figs S4} and \textbf{S5} demonstrate a high-scoring diagram, ground truth and generated, respectively. \textbf{Figs S6} and \textbf{S7} demonstrate a low-scoring diagram, ground truth and generated, respectively. A high SSIM and BLEU score signifies that the generated UML diagram aligns visually and textually with the reference, preserving spatial arrangement and semantic content. Conversely, low scores indicate discrepancies, such as misaligned components or incorrect labels, compromising structural fidelity. These examples underscore the critical role of SSIM and BLEU in validating the performance of models tasked with UML code generation. Maintaining high visual and textual accuracy is essential to ensure the diagrams remain functional for software development workflows.

Several metrics were incorporated to evaluate our models' computational efficiency, including Floating-Point Operations (FLOS), training time, samples per second, steps per second, and evaluation time. FLOS represents the cumulative number of operations during training, offering a measure of computational demand. As the model size and dataset complexity increase, FLOS and training time also rise, revealing scalability challenges. Samples per second and steps per second measure throughput and gradient updates, respectively, providing insights into how efficiently the models handle large datasets. Evaluation time measures the latency of generating a response to input image-prompt pairs, with longer times highlighting potential bottlenecks in handling more complex UML diagrams.

The error analysis follows a structured evaluation process inspired by methodologies from prior studies \citep{conrardy2024}. Three main error types are examined: syntax errors, UML code absence, and diagram mismatches. Syntax errors refer to missing symbols or incomplete code blocks that prevent the diagram from rendering correctly. These errors are flagged automatically when the PlantUML rendering tool fails to generate valid diagrams. UML absence errors occur when the generated output is irrelevant or incomplete and contains placeholder text or non-UML responses. The diagram mismatch score quantifies logical inconsistencies, such as producing an incorrect diagram type (e.g., a class diagram instead of a sequence diagram). These error metrics comprehensively evaluate the model's reliability, with lower scores indicating better performance.

The combined use of BLEU, SSIM, and error metrics provides robust diagram-to-code generation models' accuracy of UML diagram-to-code generation models. The study analyzes cost and output quality by analyzing these metrics across data fine-tuning approaches to identify the optimal balance between computational cost and output quality, as well as fine-tuning approaches. Structured evaluation ensures that the textual and visual components align with the ground truth, while error analysis highlights areas that require further refinement.

\section{Results}

This section evaluates how the LLaVA-1.5 models perform across various datasets and scenarios, focusing on both computational efficiency and the quality of the generated diagrams compared to the ground truth. The computational analysis considers the total time spent and the overall training loss. The methodology states that BLUE and SSIM are used to assess model performance. These metrics provide insight into how well the models generate diagrams, examining aspects such as fluency, accuracy, and structural similarity to their original counterparts. Beyond performance evaluations, an error analysis was performed on the types of mistakes that emerge in diagram generation. Error analysis focuses on syntax issues, the absence of UML code, and mismatching diagram types.

We propose an experimental hypothesis to investigate the impact of model size, dataset scale, and fine-tuning techniques on the accuracy and efficiency of UML diagram-to-code generation. Building on the overarching hypothesis outlined in Section 1, which is that fine-tuned models will outperform the baseline, the experiments aim to assess how these factors influence the performance of BLEU and SSIM metrics. We hypothesize that increasing model and dataset sizes, from small to extra-large, will significantly enhance performance, resulting in greater textual and visual fidelity, fewer syntax errors, and improved alignment with ground truth diagrams. Furthermore, full and LoRA-based fine-tuning is expected to reduce errors such as syntax issues, UML omissions, and diagram mismatches while ensuring stronger adherence to the original UML structure. 

LoRA fine-tuning results are anticipated to be comparable to full fine-tuning, with the added benefit of computational efficiency. Additionally, we predict that sequence diagrams, due to their more straightforward structure, will yield higher BLEU scores than activity diagrams. However, given the inherent complexity of activity diagrams, the latter is expected to benefit more significantly from fine-tuning. Finally, we hypothesize that expanding the dataset size will enhance the models' scalability and generalization, driving improvements across all major performance metrics. This hypothesis builds upon and complements the overarching premise, serving as the foundation for the experimental evaluation presented in the subsequent sections.
\subsection{Model Performance}

\begin{figure}[htbp]
    \centering
    \includegraphics[width=\columnwidth]{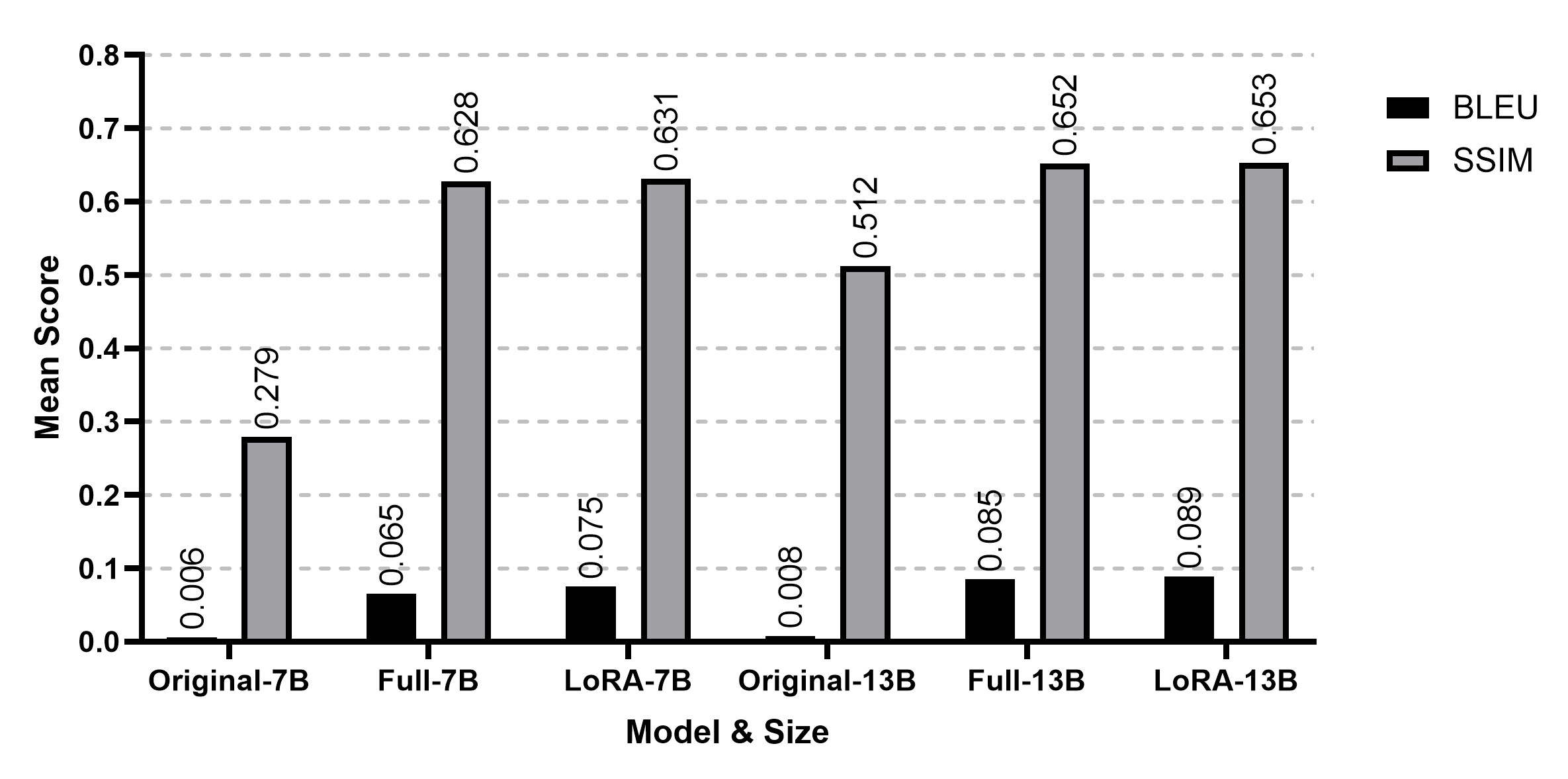}
    \caption[Small Activity Dataset Performance]{Model performance on the small activity dataset, grouped by model type and size, with BLEU and SSIM scores for each configuration.}
    \label{fig:activity_dataset_charts_small}
\end{figure}

\begin{figure}[htbp]
    \centering
    \includegraphics[width=\columnwidth]{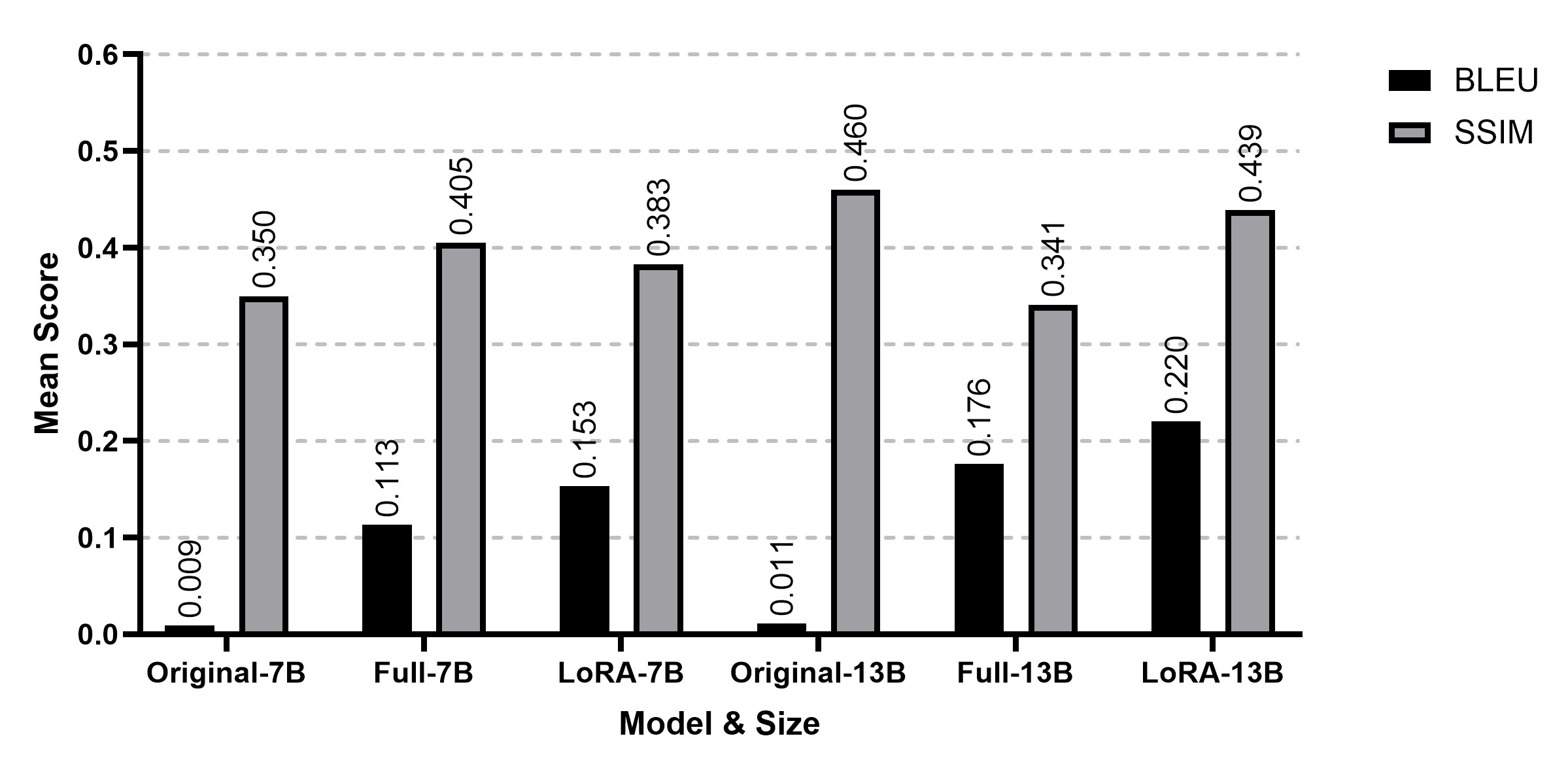}
    \caption[Small Sequence Dataset Performance]{Model performance on the small sequence dataset, grouped by model type and size, with BLEU and SSIM scores for each configuration.}
    \label{fig:sequence_dataset_charts_small}
\end{figure}

\begin{figure}[htbp]
    \centering
    \includegraphics[width=\columnwidth]{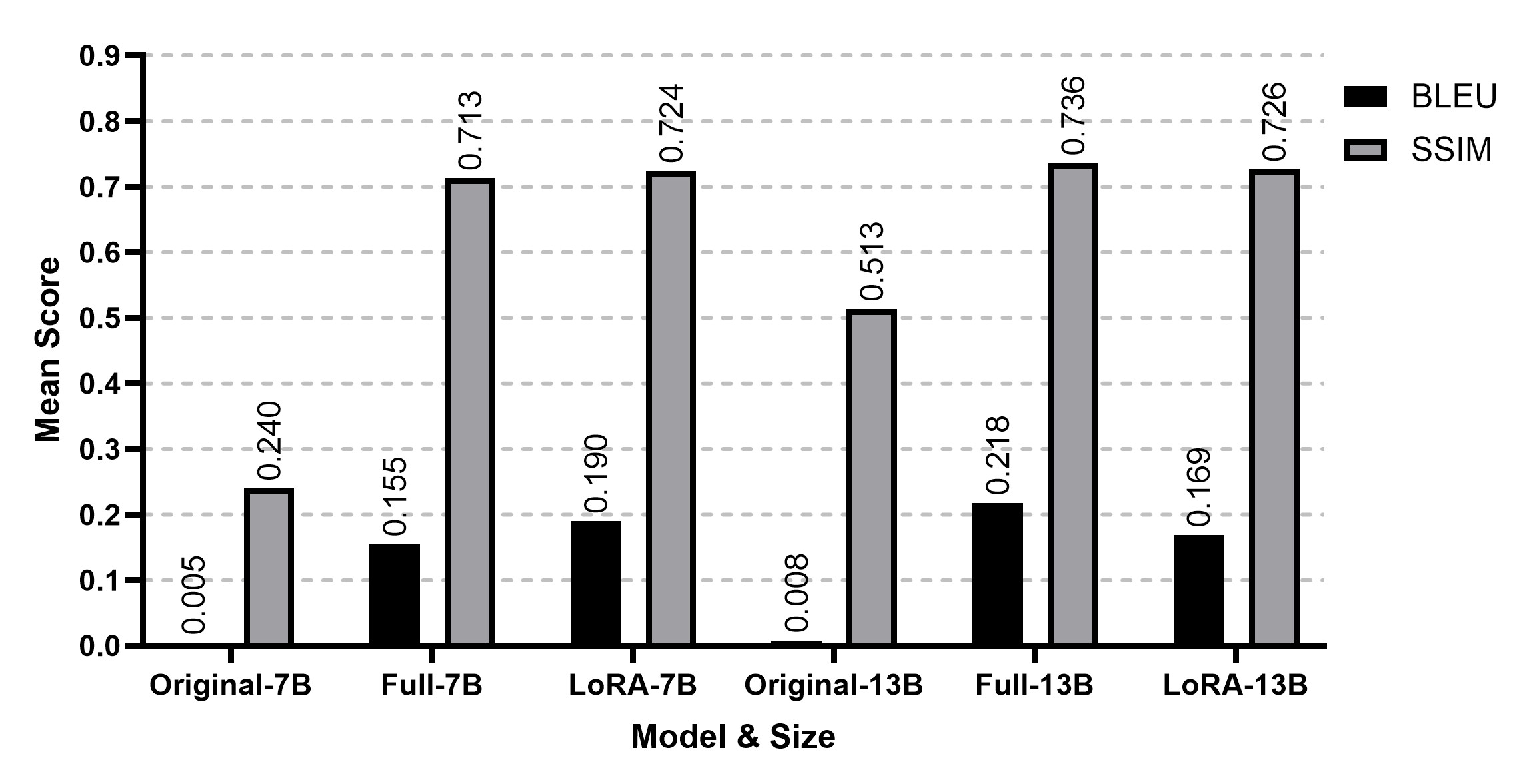}
    \caption[Medium Activity Dataset Performance]{Model performance on the medium activity dataset, grouped by model type and size, with BLEU and SSIM scores for each configuration.}
    \label{fig:activity_dataset_charts_medium}
\end{figure}

\begin{figure}[htbp]
    \centering
    \includegraphics[width=\columnwidth]{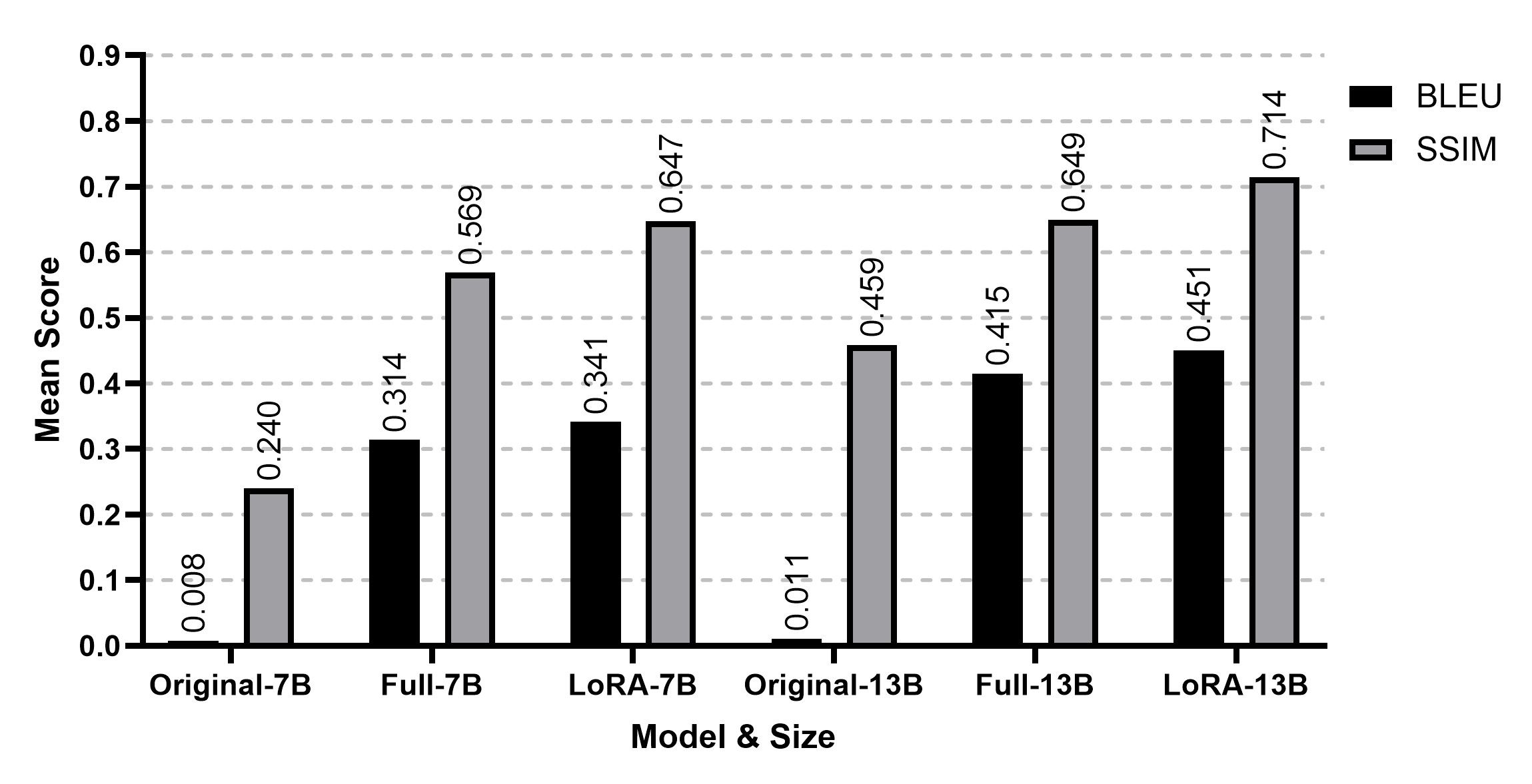}
    \caption[Medium Sequence Dataset Performance]{Model performance on the medium sequence dataset, grouped by model type and size, with BLEU and SSIM scores for each configuration.}
    \label{fig:sequence_dataset_charts_medium}
\end{figure}

\begin{figure}[htbp]
    \centering
    \includegraphics[width=\columnwidth]{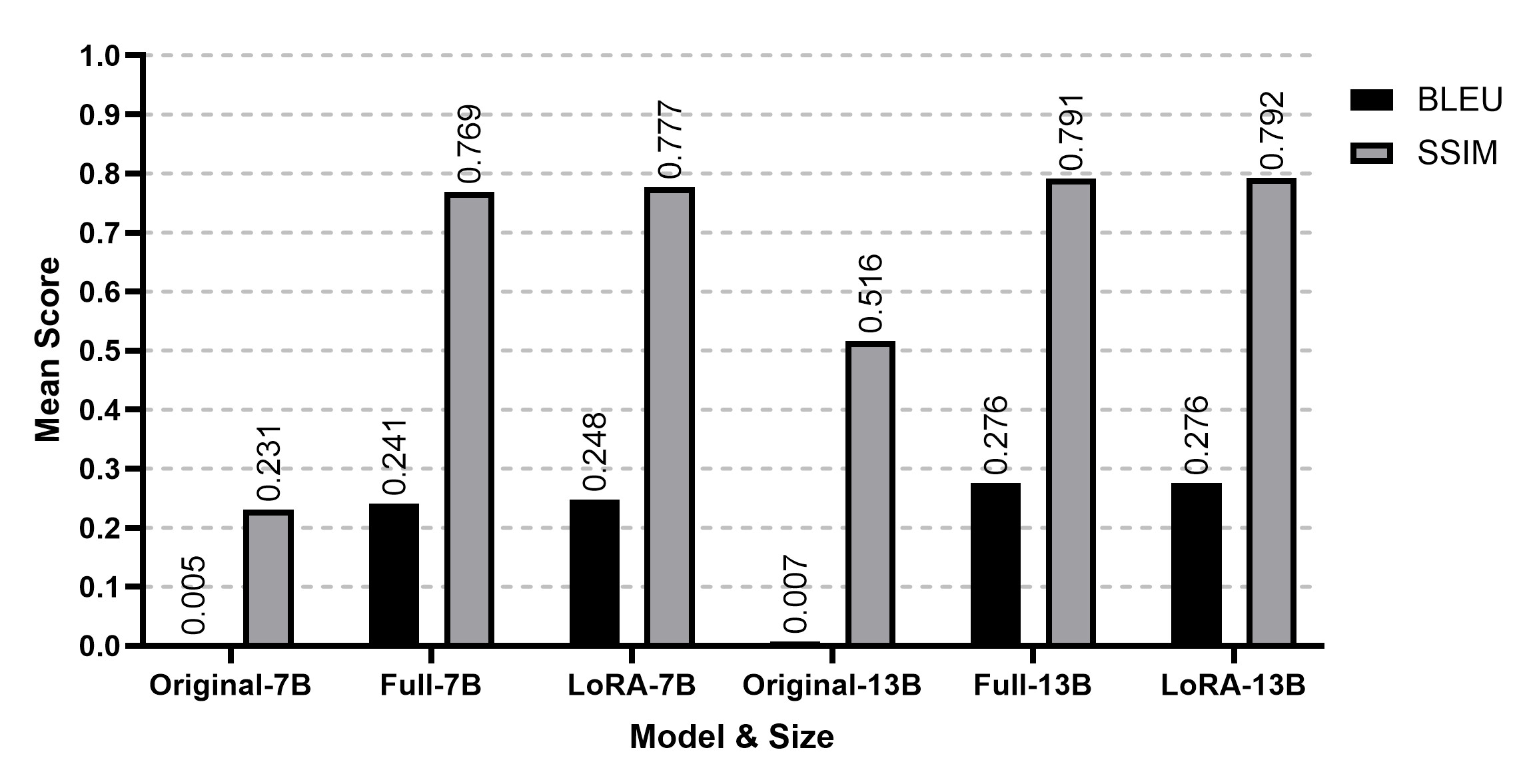}
    \caption[Large Activity Dataset Performance]{Model performance on the large activity dataset, grouped by model type and size, with BLEU and SSIM scores for each configuration.}
    \label{fig:activity_dataset_charts_large}
\end{figure}

\begin{figure}[htbp]
    \centering
    \includegraphics[width=\columnwidth]{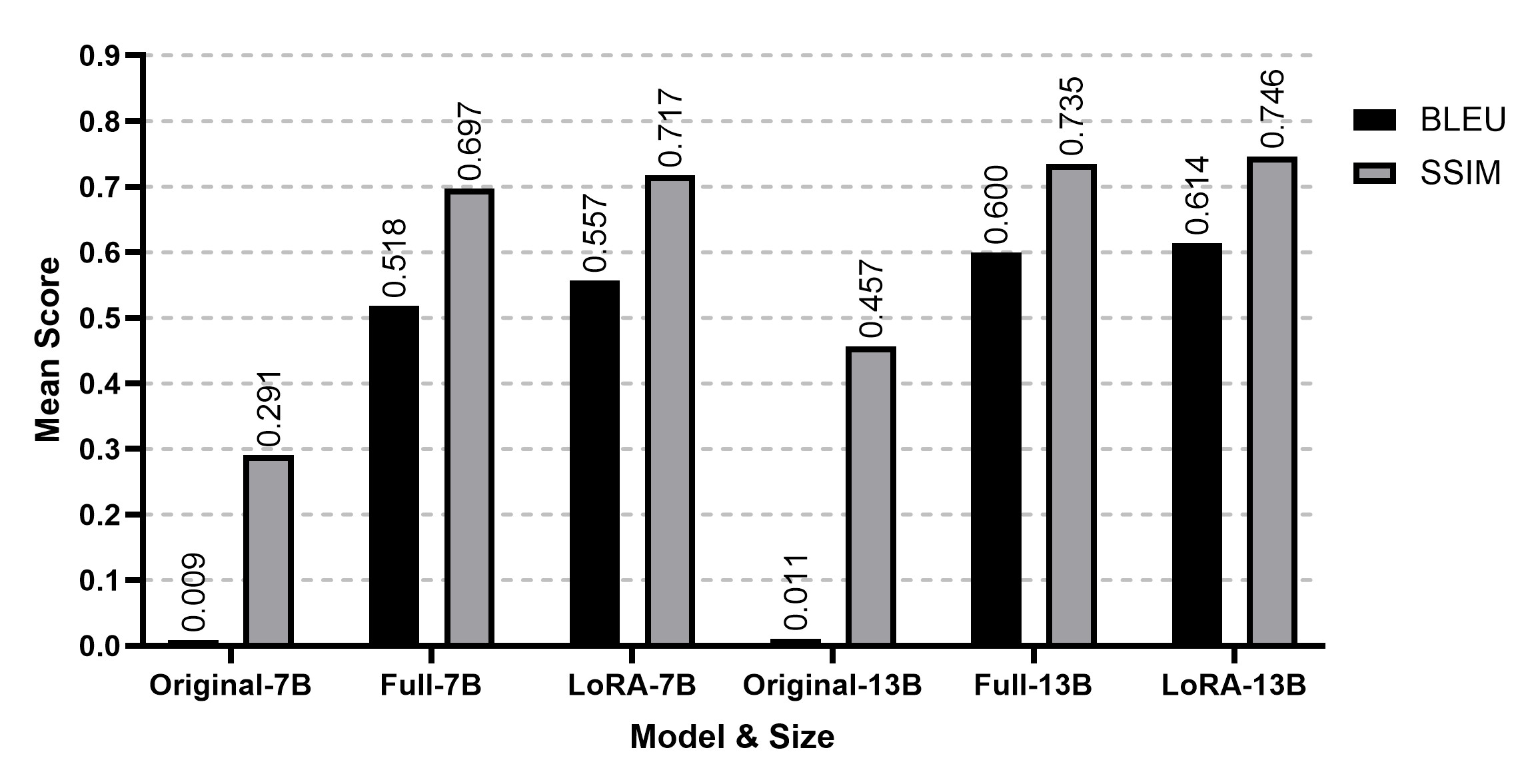}
    \caption[Large Sequence Dataset Performance]{Model performance on the large sequence dataset, grouped by model type and size, with BLEU and SSIM scores for each configuration.}
    \label{fig:sequence_dataset_charts_large}
\end{figure}

\begin{figure}[htbp]
    \centering
    \includegraphics[width=\columnwidth]{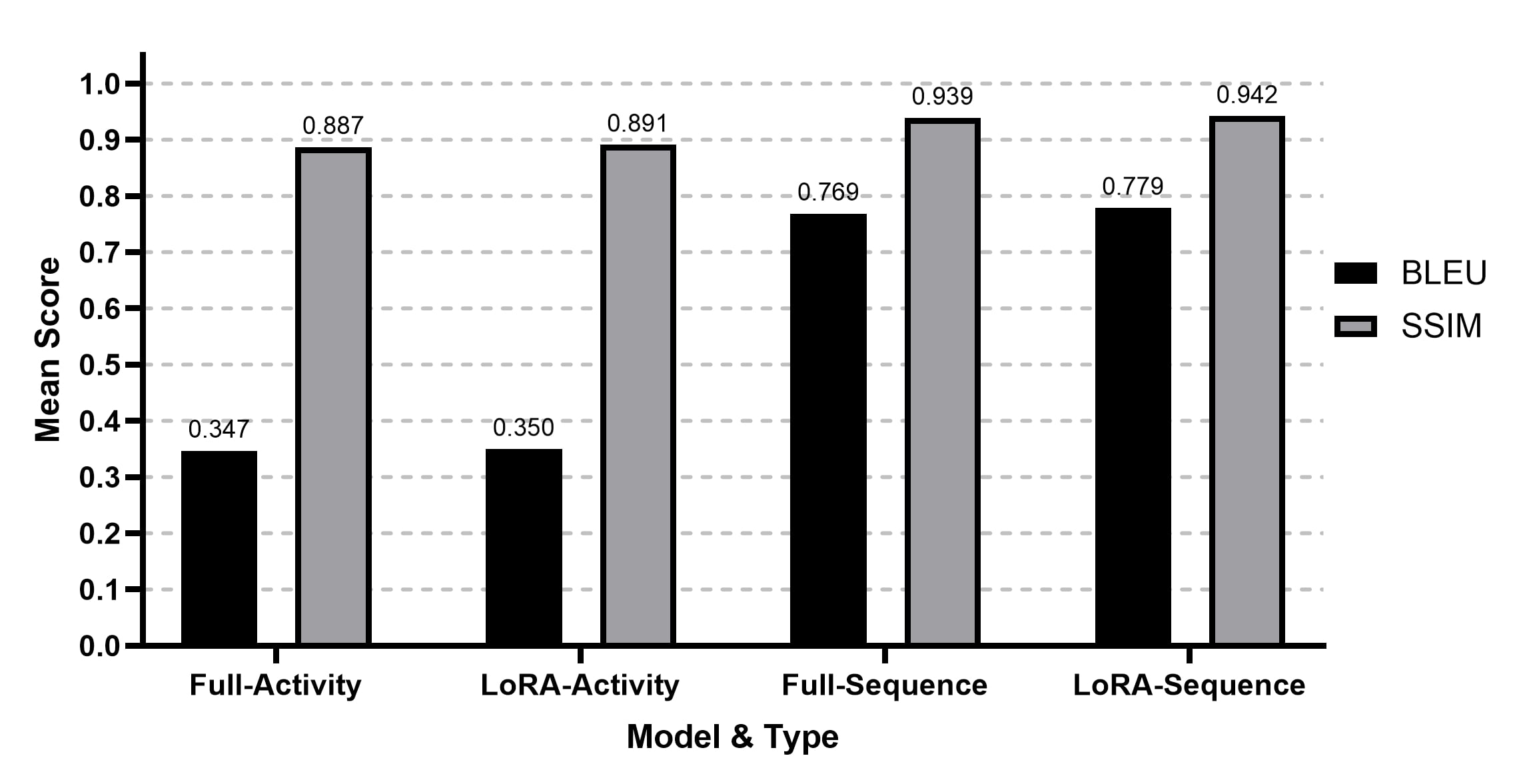}
    \caption[Extra Large Dataset Performance]{Model performance on the extra-large dataset for activity and sequence diagrams, grouped by BLEU and SSIM scores across different model types and data types.}
    \label{fig:activity_and_sequence_dataset_charts_extra_large}
\end{figure}

BLEU scores tend to be lower at small dataset sizes. As seen in Fig \ref{fig:sequence_dataset_charts_small}, the original 7B model’s BLEU score was only 0.009 for the small sequence dataset, which is notably lower than both the full 7B model (0.113) and the LoRA 7B model (0.153). The original 13B model also performed poorly in the same small sequence dataset, registering a BLEU score of 0.011. A similar pattern appears in the small activity dataset, seen in Fig \ref{fig:activity_dataset_charts_small}, where low BLEU scores underscore the challenge of generating accurate outputs from such limited data. Nonetheless, even minimal fine-tuning appears to be more effective than the original model at recreating UML diagrams.
As the dataset size increases, the same general trends emerge, as shown in Figs  \ref{fig:activity_dataset_charts_medium}, \ref{fig:sequence_dataset_charts_medium}, \ref{fig:activity_dataset_charts_large}, and \ref{fig:sequence_dataset_charts_large}. LoRA models usually outperform—or at least match—the full models, while the baseline tends to perform the worst. One exception is that the original 13B model outperformed the full 13B sequence data model. Moving on to the extra-large dataset, seen in Fig \ref{fig:activity_and_sequence_dataset_charts_extra_large}, the LoRA 13B model achieved BLEU scores of 0.779 for sequence diagrams and 0.350 for activity diagrams, far surpassing earlier results. The full 13B model also showed comparable gains, scoring 0.769 and 0.347, respectively. The higher scores illustrate the significant benefits of using larger datasets.

SSIM scores followed a similar trajectory. In the small activity dataset, the baseline 7B model only reached 0.279, whereas the full and LoRA 7B models exceeded 0.600. Although the 7B model performed slightly better on the small sequence dataset, sequence SSIM scores remained below most of the activity. The lower scores are not seen the extra large dataset where SSIM scores outperform activity. With medium and large datasets, SSIM values continued to rise. Finally, the extra-large dataset delivered the highest SSIM scores overall. As seen in Fig \ref{fig:activity_and_sequence_dataset_charts_extra_large}, the LoRA 13B model achieved 0.942 on sequence diagrams and 0.891 on activity diagrams, with the full 13B model closely matching these values. Such results suggest that larger datasets and fine-tuning strategies enable models to capture structural details more effectively, resulting in higher BLEU scores and stronger SSIM performance.

Across all dataset sizes, models trained on sequence data consistently outperformed those trained on activity data, achieving higher BLEU and SSIM scores. The LoRA 13B model achieved the highest scores, with 0.779 for BLEU and 0.942 for SSIM on sequence data in the extra-large dataset seen in Fig \ref{fig:activity_and_sequence_dataset_charts_extra_large}. This discrepancy reflects the relative simplicity of sequence data compared to activity diagrams, which involve more intricate structural and syntactic complexity. While LoRA fine-tuning consistently enhanced performance across metrics, its benefits were more pronounced in sequence datasets. The performance gap between full and LoRA models was smaller for activity diagrams, suggesting that LoRA's targeted parameter updates are particularly effective for less complex data. These findings reinforce the critical role of fine-tuning in improving performance across different UML diagram types and dataset sizes.

\subsubsection{Real-World Evaluation and Limitations}

\begin{figure}[htbp]
    \centering
    \includegraphics[width=\columnwidth]{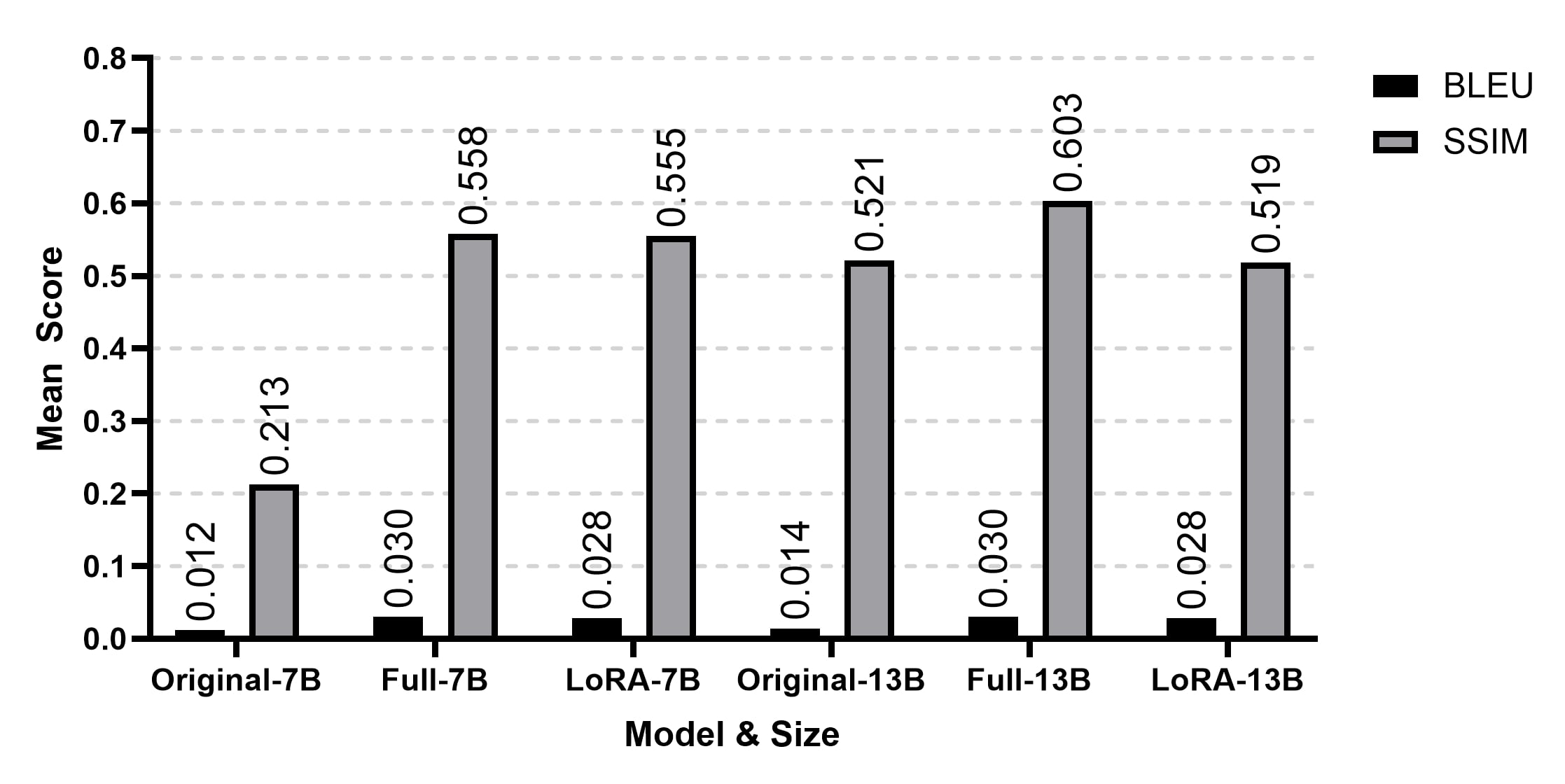}
    \caption[Real-World Activity Diagram Performance]{Model performance on the real-world activity diagram dataset, showing BLEU and SSIM scores across baseline, full fine-tuning, and LoRA configurations with the small dataset models. Compared to synthetic data, improvements from fine-tuning are more modest.}
    \label{fig:realworld_activity_performance}
\end{figure}

\begin{figure}[htbp]
    \centering
    \includegraphics[width=\columnwidth]{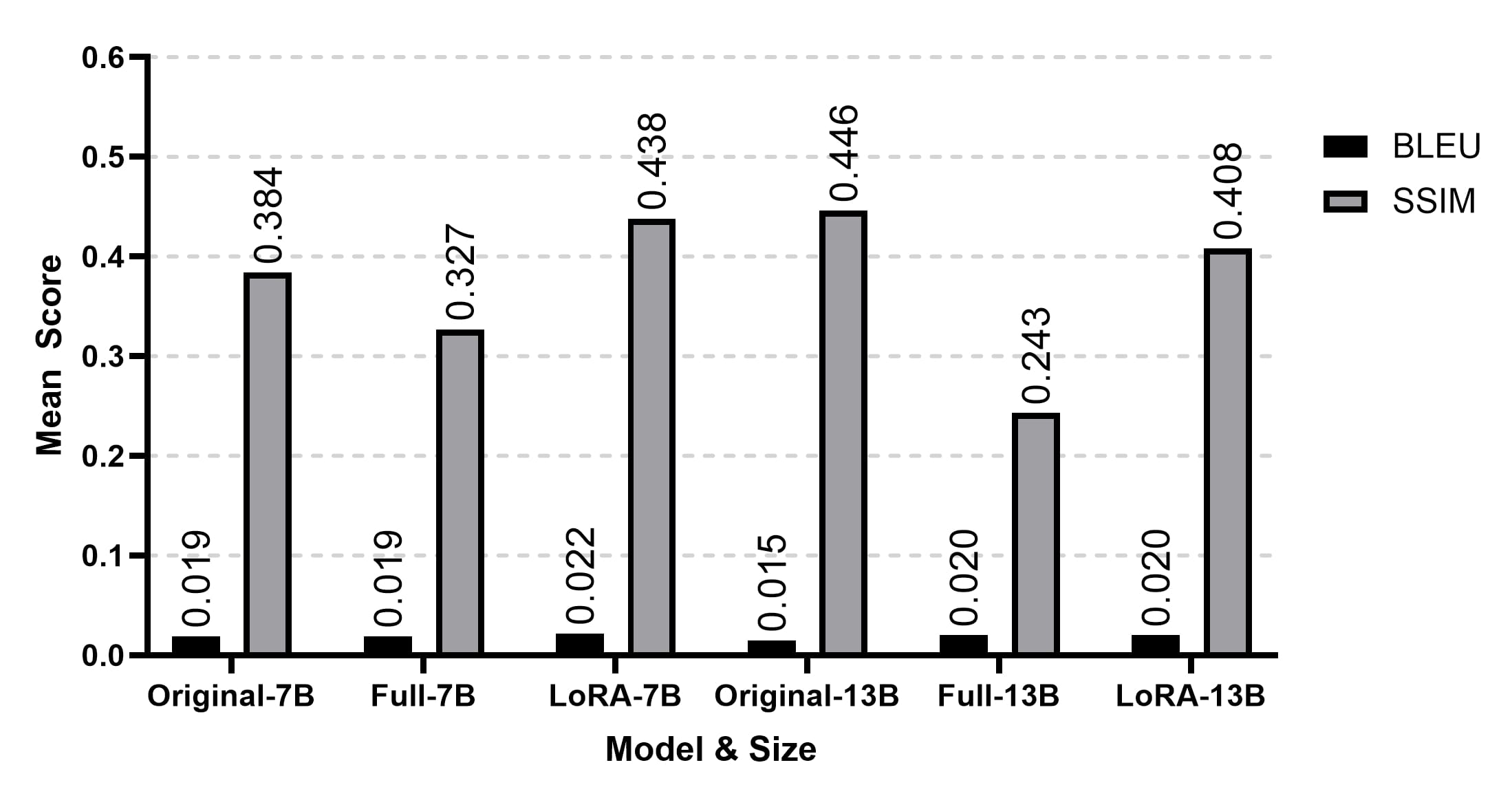}
    \caption[Real-World Sequence Diagram Performance]{Model performance on the real-world  sequence diagram dataset. Fine-tuned models, for the small dataset models,  show only minor gains over the baseline, suggesting that visual mismatches between training and test data affect generalization.}
    \label{fig:realworld_sequence_performance}
\end{figure}

\begin{figure}[htbp]
    \centering
    \includegraphics[width=\columnwidth]{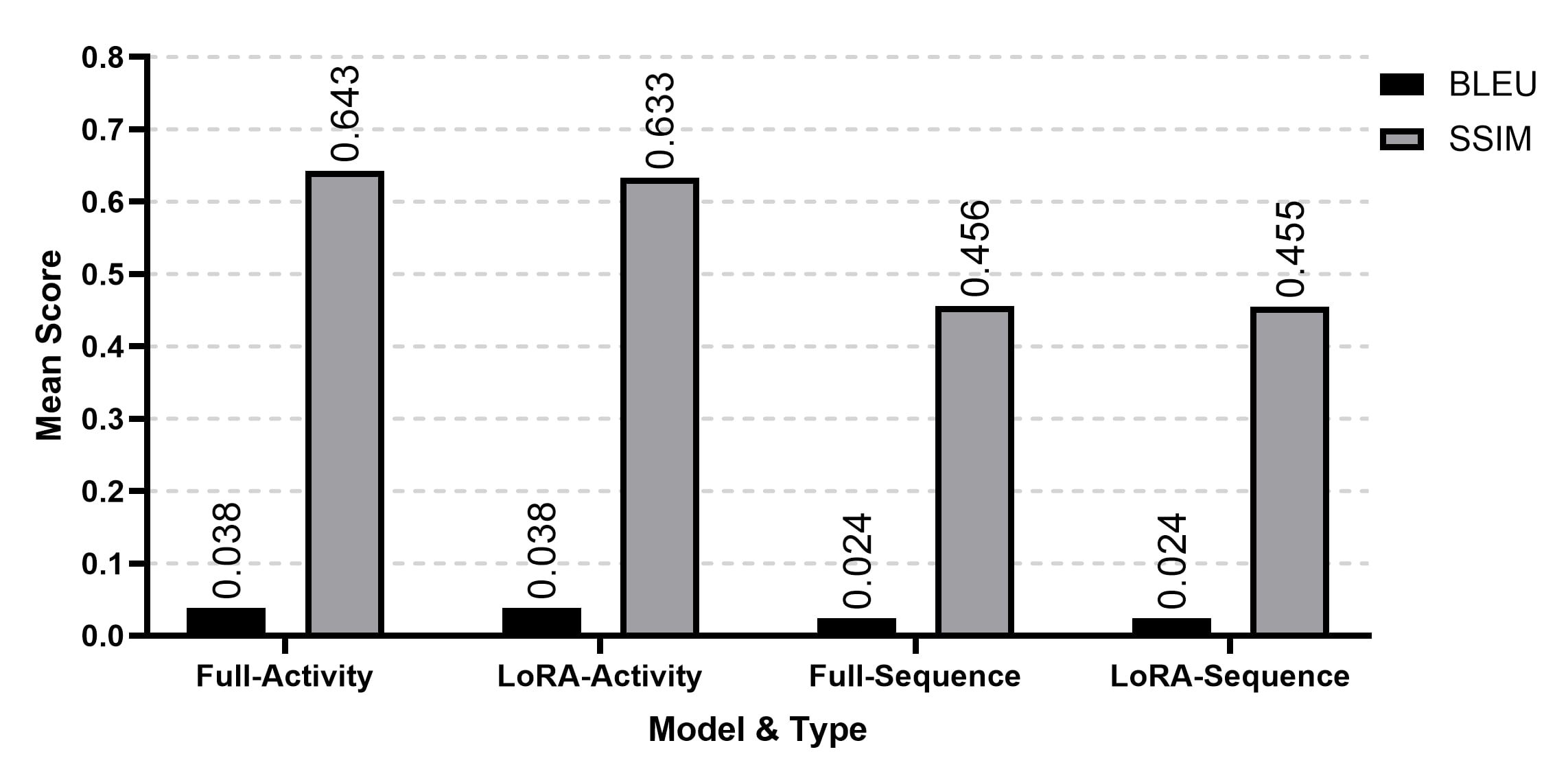}
    \caption[Combined Real-World UML Performance]{Comparison of BLEU and SSIM scores across real-world activity and sequence diagrams. Results indicate that both full and LoRA fine-tuning yield limited improvements over the baseline, underscoring the importance of aligning synthetic training data with real-world visual styles.}
    \label{fig:realworld_combined_performance}
\end{figure}

Performance on real-world diagrams remained limited, as shown in Figures \ref{fig:realworld_activity_performance}, \ref{fig:realworld_sequence_performance}, and \ref{fig:realworld_combined_performance}. While increasing model size and training duration led to slight gains—BLEU scores for the activity test set improved from 0.012 of the orignial 7B model to 0.038 for the extra large LoRA training data model, and SSIM increased from 0.213 to 0.643—these improvements were modest. Results showed an even smaller increase in sequence data from the small training models to the extra large training models. The relatively small improvements underscore the gap between training conditions and real-world application.

Supplementary figures, \textbf{Fig. S8} through \textbf{Fig. S11}, visualize these limitations, contrasting model-generated outputs with actual UML diagrams. The differences in terminology, formatting, and syntactic structure make it clear that synthetic data lacks realism in key areas. These observations point to the need for more representative training data that better mirrors real-world patterns. Given the difficulty of scaling real-world data collection, a practical path forward involves refining synthetic data generation to better simulate real-world use. Designing datasets around specific application domains—such as authentication, e-commerce, or user management—and adopting realistic naming conventions could significantly enhance generalization.

\subsection{Computational Efficiency}

\subsubsection{Training Trends}

The computational demands of machine learning model training depend on the size of the dataset, the complexity of the dataset, and the model architecture. Tables \ref{tab:training_time_activity_7B}, \ref{tab:training_time_activity_13B}, \ref{tab:training_time_sequence_7B}, and \ref{tab:training_time_sequence_13B} detail the metrics for training time and computational performance across activity and sequence datasets of varying sizes (small to extra-large) for both the 7B and 13B model architectures. While activity and sequence datasets share trends in step rates and throughput, differences in complexity lead to notable disparities in computational demands. Step rates decrease slightly as the dataset size grows. For instance, the 7B full model processes 0.081, 0.080, and 0.079 steps per second for the small, medium, and large activity datasets. Small reductions in step rates reflect the increasing computational demands of larger datasets while demonstrating the model’s consistency. Similarly, the 7B full model maintains a steady 0.108 steps per second across small, medium, and large sequence datasets, highlighting the lighter computational burden of sequence data. These results confirm the robustness of the model architecture in maintaining efficiency across different scales and types of datasets.

\begin{table*}[!ht] 
\centering
\scriptsize
\caption{Training Time Metrics for Activity Datasets by Model Size: 7B Model}
\label{tab:training_time_activity_7B}
\resizebox{\linewidth}{!}{%
\begin{tabular}{l c c c c c c}
\hline
\textbf{Dataset Size} & \textbf{Model Type} & \textbf{Training Time (hrs)} & \textbf{Train Samples Per sec} & \textbf{Train Steps Per sec} & \textbf{Train Loss} & \textbf{Total FLOS} \\ \hline
Small       & Full  & 0.161 & 10.335 & 0.081 & 2.020 & $8.721 \times 10^{13}$ \\ 
Small       & LoRA  & 0.252 & 6.626  & 0.052 & 2.022 & $9.731 \times 10^{13}$ \\ 
Medium      & Full  & 0.323 & 10.305 & 0.080 & 1.924 & $1.857 \times 10^{14}$ \\ 
Medium      & LoRA  & 0.507 & 6.578  & 0.051 & 1.924 & $2.066 \times 10^{14}$ \\ 
Large       & Full  & 0.658 & 10.127 & 0.079 & 1.830 & $3.880 \times 10^{14}$ \\ 
Large       & LoRA  & 1.038 & 6.423  & 0.050 & 1.800 & $4.326 \times 10^{14}$ \\ \hline
\end{tabular}%
}
\end{table*}

\begin{table*}[!ht] 
\centering
\scriptsize
\caption{Training Time Metrics for Activity Datasets by Model Size: 13B Model}
\label{tab:training_time_activity_13B}
\resizebox{\linewidth}{!}{%
\begin{tabular}{l c c c c c c}
\hline
\textbf{Dataset Size} & \textbf{Model Type} & \textbf{Training Time (hrs)} & \textbf{Train Samples Per sec} & \textbf{Train Steps Per sec} & \textbf{Train Loss} & \textbf{Total FLOS} \\ \hline
Small       & Full  & 0.284 & 5.862  & 0.046 & 1.945 & $1.090 \times 10^{14}$ \\ 
Small       & LoRA  & 0.432 & 3.858  & 0.030 & 1.953 & $1.217 \times 10^{14}$ \\ 
Medium      & Full  & 0.577 & 5.774  & 0.045 & 1.867 & $2.321 \times 10^{14}$ \\ 
Medium      & LoRA  & 0.891 & 3.743  & 0.029 & 1.871 & $2.583 \times 10^{14}$ \\ 
Large       & Full  & 1.180 & 5.650  & 0.044 & 1.777 & $4.851 \times 10^{14}$ \\ 
Large       & LoRA  & 1.784 & 3.736  & 0.029 & 1.769 & $5.408 \times 10^{14}$ \\ 
Extra Large & Full  & 5.948 & 5.604  & 0.044 & 1.569 & $2.484 \times 10^{15}$ \\ 
Extra Large & LoRA  & 9.002 & 3.703  & 0.029 & 1.553 & $2.772 \times 10^{15}$ \\ \hline
\end{tabular}%
}
\end{table*}

\begin{table*}[!ht] 
\centering
\scriptsize
\caption{Training Time Metrics for Sequence Datasets by Model Size: 7B Model}
\label{tab:training_time_sequence_7B}
\resizebox{\linewidth}{!}{%
\begin{tabular}{l c c c c c c}
\hline
\textbf{Dataset Size} & \textbf{Model Type} & \textbf{Training Time (hrs)} & \textbf{Train Samples Per sec} & \textbf{Train Steps Per sec} & \textbf{Train Loss} & \textbf{Total FLOS} \\ \hline
Small       & Full  & 0.121 & 13.738 & 0.108 & 0.615 & $4.003 \times 10^{13}$ \\ 
Small       & LoRA  & 0.190 & 8.758  & 0.069 & 0.574 & $9.731 \times 10^{13}$ \\ 
Medium      & Full  & 0.240 & 13.864 & 0.108 & 0.502 & $8.121 \times 10^{13}$ \\ 
Medium      & LoRA  & 0.374 & 8.897  & 0.069 & 0.474 & $8.545 \times 10^{13}$ \\ 
Large       & Full  & 0.483 & 13.802 & 0.108 & 0.380 & $1.699 \times 10^{14}$ \\ 
Large       & LoRA  & 0.753 & 8.852  & 0.069 & 0.357 & $1.777 \times 10^{14}$ \\ \hline
\end{tabular}%
}
\end{table*}

\begin{table*}[!ht] 
\centering
\scriptsize
\caption{Training Time Metrics for Sequence Datasets by Model Size: 13B Model}
\label{tab:training_time_sequence_13B}
\resizebox{\linewidth}{!}{%
\begin{tabular}{l c c c c c c}
\hline
\textbf{Dataset Size} & \textbf{Model Type} & \textbf{Training Time (hrs)} & \textbf{Train Samples Per sec} & \textbf{Train Steps Per sec} & \textbf{Train Loss} & \textbf{Total FLOS} \\ \hline
Small       & Full  & 0.214 & 7.805  & 0.061 & 0.544 & $5.005 \times 10^{13}$ \\ 
Small       & LoRA  & 0.334 & 4.995  & 0.039 & 0.515 & $5.285 \times 10^{13}$ \\ 
Medium      & Full  & 0.427 & 7.801  & 0.060 & 0.433 & $1.015 \times 10^{14}$ \\ 
Medium      & LoRA  & 0.642 & 5.189  & 0.040 & 0.417 & $1.068 \times 10^{14}$ \\ 
Large       & Full  & 0.950 & 7.819  & 0.061 & 0.324 & $2.124 \times 10^{14}$ \\ 
Large       & LoRA  & 1.313 & 5.075  & 0.040 & 0.314 & $2.222 \times 10^{14}$ \\ 
Extra Large & Full  & 4.290 & 7.770  & 0.061 & 0.178 & $1.075 \times 10^{15}$ \\ 
Extra Large & LoRA  & 6.514 & 5.117  & 0.040 & 0.171 & $1.123 \times 10^{15}$ \\ \hline
\end{tabular}%
}
\end{table*}

Activity datasets typically exhibit greater complexity, resulting in slower step rates than sequence datasets. For example, the 7B full model processes 10.335 samples per second on the small activity dataset versus 13.738 samples per second on the corresponding sequence dataset. Activity datasets require more operations per sample due to their richer feature sets, which slows processing rates and lowers throughput. However, the consistent step rates within each dataset type indicate that the architecture handles increasing dataset size without introducing performance bottlenecks.

Training time increases significantly with dataset size for both activity and sequence datasets, though activity datasets demand more time due to their complexity. For instance, training the 7B full model on the small activity dataset takes 0.161 hours, compared to 0.121 hours for the equivalent sequence dataset. This trend holds across all dataset sizes, with the 13B full model requiring 5.948 hours for the extra-large activity dataset and 4.290 hours for the corresponding sequence dataset. The additional time required for activity datasets stems from their slower step rates and lower throughput. For example, the 7B full model processes 10.335 samples per second for the small activity dataset but achieves 13.738 samples per second for the small sequence dataset. Despite these differences, the processing consistency within each dataset type showcases the scalability of the 7B and 13B models. 

Activity datasets consistently require higher FLOS than sequence datasets of the same size. For instance, training the 7B full model on the small activity dataset involves $8.721 \times 10^{13}$ FLOS, compared to $4.003 \times 10^{13}$ FLOS for the equivalent sequence dataset. This disparity becomes more pronounced with larger datasets, as the 13B full model demands $2.484 \times 10^{15}$ FLOS for the extra-large activity dataset but only $1.075 \times 10^{15}$ FLOS for the sequence dataset. Sequence datasets are less computationally intensive, supporting faster convergence and reduced training time, making them advantageous for applications requiring quicker deployment.
Training loss trends reveal that activity and sequence datasets yield lower loss values as dataset and model sizes grow. For instance, the 7B full model achieves a loss of 2.020 on the small activity dataset, while the equivalent sequence dataset achieves a lower loss of 0.615. Larger models and datasets amplify this pattern, with the 13B full model attaining a loss of 1.553 for the extra-large activity dataset and 0.178 for the corresponding sequence dataset. The lower loss values associated with sequence datasets reflect more efficient learning and potentially better generalization. Conversely, the higher loss values for activity datasets may challenge optimization but offer richer feature spaces. These features could capture more nuanced patterns, benefiting specific applications despite higher computational costs.

\subsubsection{Evaluation Time}

\begin{figure}[htbp]
    \centering
    \includegraphics[width=\columnwidth]{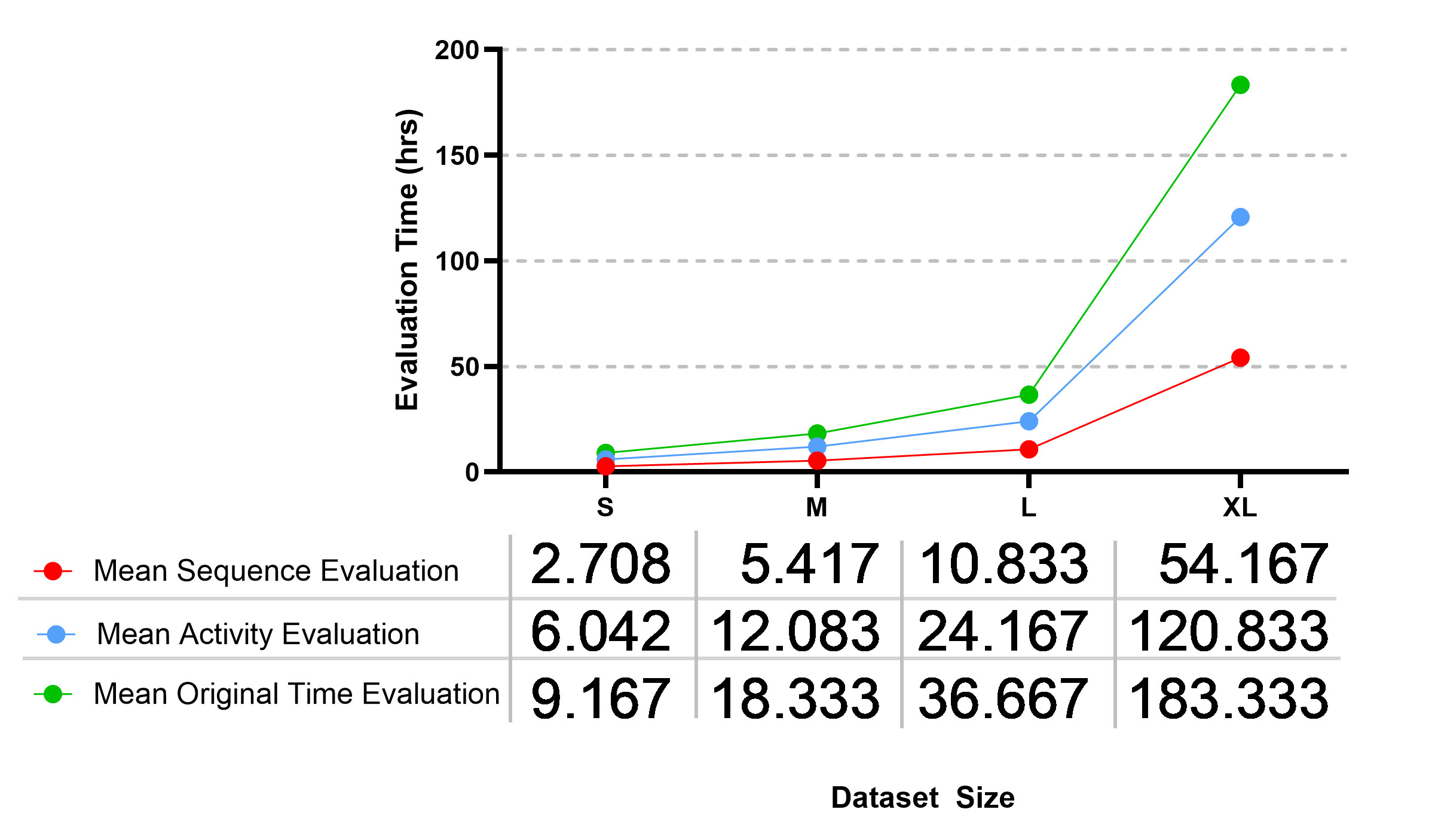}
    \caption[Evaluation Time]{Evaluation time across various model configurations and dataset conditions, illustrating different setups' computational efficiency and response behavior.}
    \label{fig:evaluation_time}
\end{figure}

Fig \(\ref{fig:evaluation_time}\) illustrates the evaluation time for various models, including a non-fine-tuned baseline. Evaluation time primarily depends on the model's response length to image-prompt pairs and the total number of pairs rather than the inherent complexity of the data. For sequence models, evaluation time scales with dataset size, increasing from 2.708 hours for the small dataset to 54.167 hours for the extra-large dataset. This increase stems from the larger number of image-prompt pairs, requiring the model to generate more responses. The relatively straightforward structure of sequence diagrams results in concise outputs, helping to keep evaluation times relatively lower compared to activity models.
In contrast, models fine-tuned on activity data demonstrate longer evaluation times due to the intricate nature of activity diagrams, which necessitate more detailed and extended responses. Evaluation time begins at 6.042 hours for the small dataset and rises to 120.833 hours for the extra-large dataset. The gap between sequence and activity models widens as dataset size increases, highlighting the greater computational demands of processing more complex diagram types. 

The non-fine-tuned baseline model shows the longest evaluation times across all dataset sizes, starting at 9.167 hours for the small dataset and reaching 183.333 hours for the extra-large dataset. This model generates verbose and often inaccurate responses, extending evaluation time beyond fine-tuned models for sequence and activity data. These findings emphasize that evaluation time, rather than training time, becomes the primary bottleneck when working with large datasets. Optimizing evaluation processes—such as reducing unnecessary response lengths—is crucial for efficiently deploying models. Fine-tuning not only improves response accuracy but also curtails evaluation time.

\subsection{Error Analysis}

The original 7B model exhibits the highest syntax error rates when generating sequence diagrams, particularly for small and medium datasets. Specifically, it produces rates of 0.343 for the small dataset and 0.522 for the medium dataset before stabilizing at 0.439 for larger datasets. In contrast, the full and LoRA-tuned versions of the 7B model show marked improvements, with the LoRA variant reducing syntax errors to 0.297. The 13B models follow a similar pattern but with consistently lower syntax error rates. The original 13B model starts at 0.133 for small datasets, with reductions observed as dataset size increases. Notably, the 13B LoRA model achieves the lowest syntax error rate, dropping to 0.031 for large datasets and 0.006 for extra-large datasets. This highlights the benefits of larger model sizes and fine-tuning strategies to reduce syntax errors effectively.

More complex activity diagrams initially exhibit higher syntax error rates, particularly for the baseline 7B model. The syntax error rate reaches 0.623 for small datasets, increasing slightly to 0.683 for larger datasets. However, fine-tuning mitigates these errors. The full 7B model achieves a syntax error rate of 0.012, and the LoRA-tuned 7B model follows closely with 0.019 for the small dataset. The 13B LoRA model again demonstrates the best performance, achieving an error rate of 0.006 for the extra-large dataset. These results indicate that fine-tuning, particularly LoRA-based strategies, is highly effective for complex activity diagrams.

\begin{figure}[htbp]
    \centering
    \includegraphics[width=0.8\textwidth, trim=0 0 500 0, clip]{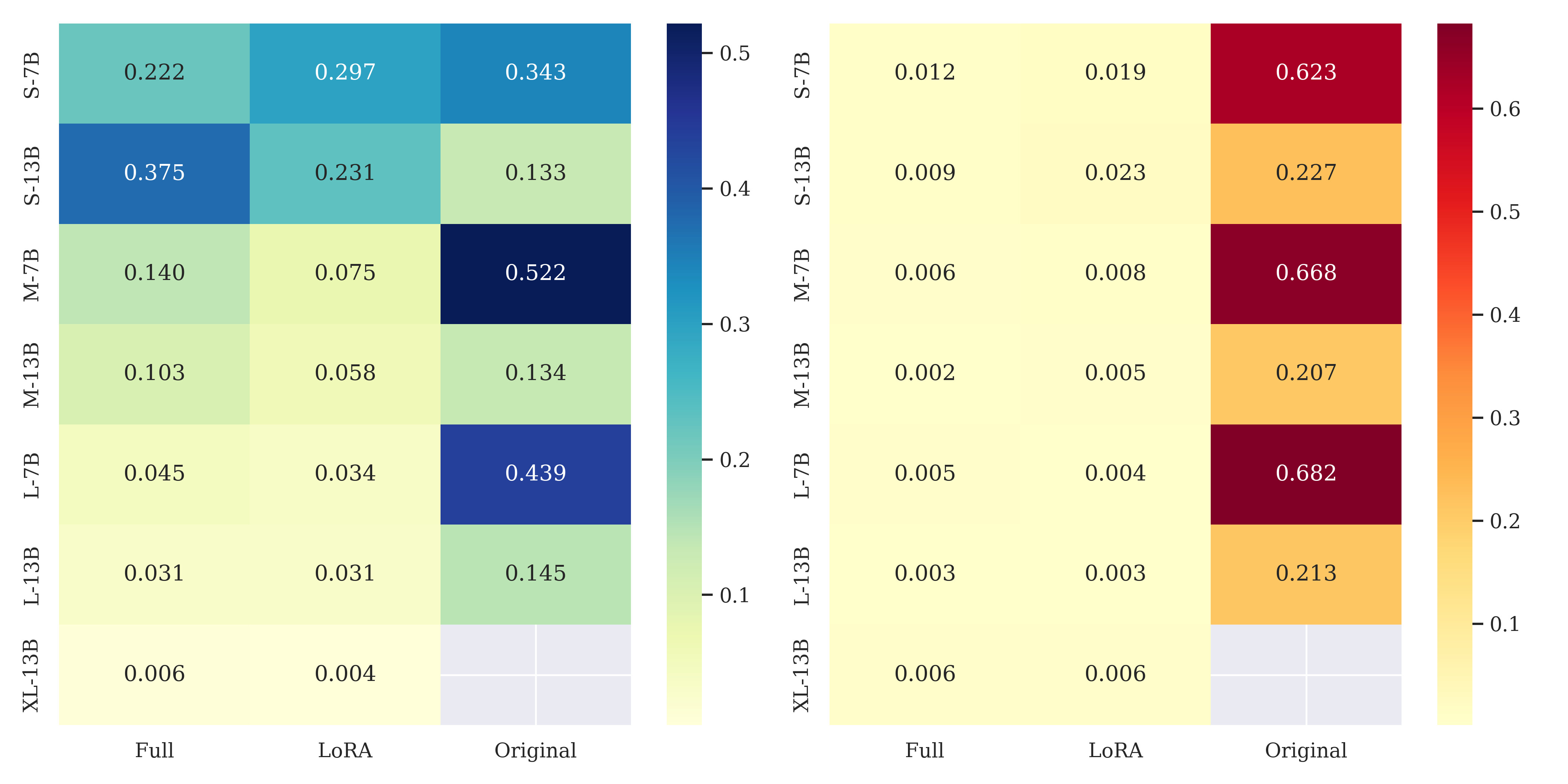}
    \caption{Syntax error rates for sequence diagrams across all dataset sizes and model configurations. Darker shades represent higher error frequencies.}
    \label{fig:syntax_error_heatmap_sequence}
\end{figure}

\begin{figure}[htbp]
    \centering
    \includegraphics[width=0.8\textwidth, trim=488 0 0 0, clip]{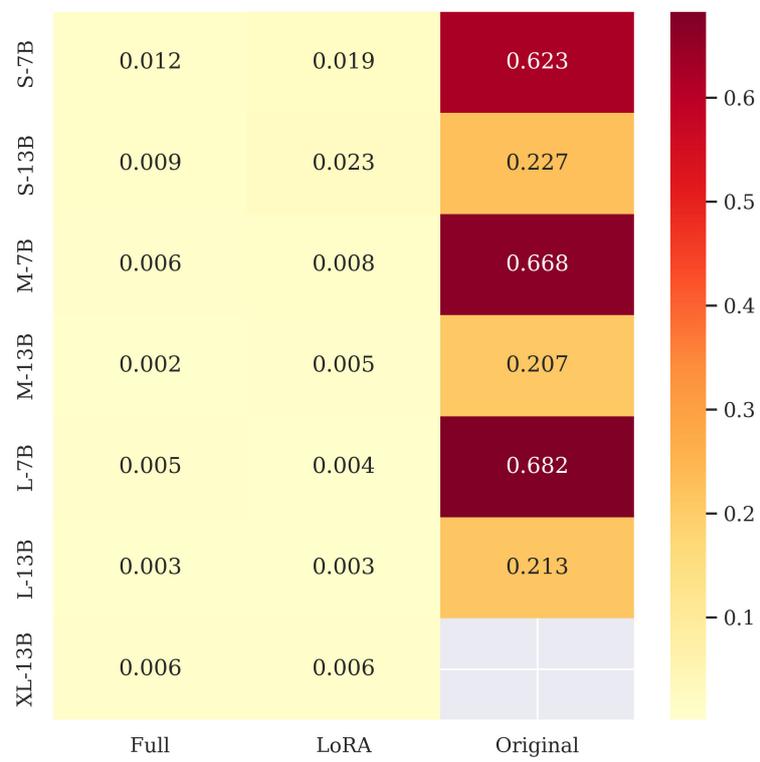}
    \caption{Syntax error rates for activity diagrams across all dataset sizes and model configurations. Darker shades represent higher error frequencies.}
    \label{fig:syntax_error_heatmap_activity}
\end{figure}

The UML absence score measures instances where models fail to generate UML code. Table\ref{tab:error_analysis} shows that baseline models display slow but non-negligible absence rates, while fine-tuned models eliminate this timely For the small dataset, the baseline 7B model records a UML absence score of 0.0007 for both sequence and activity diagrams. The baseline 13B model avoids absence errors for small datasets but introduces minor errors for medium and large datasets. A notable exception occurs with the 7B model on the medium-sized sequence dataset, where the absence score increases to 0.0023.

The diagram mismatch score measures how often the generated diagram type differs from the prompt. Baseline models again demonstrate higher mismatch rates, particularly for activity diagrams. For example, the 13B baseline model records a mismatch score of 0.75 for small datasets, rising slightly to 0.7538 for large datasets. Sequence diagrams also show problems in baseline models, where the 7B model generates mismatches at a rate of 0.3567 for small datasets and 0.5254 for medium datasets. Fine-tuned models resolve this problem entirely, producing diagrams that consistently match the requested type. The reduction in mismatch scores highlights the effectiveness of fine-tuning to adhere to prompt instructions and improve the accuracy of the model output.

\begin{table*}[htbp]
\centering
\caption{UML Absence and Diagram Mismatch Scores by Dataset Size and Model Configuration}
\label{tab:error_analysis}
\resizebox{\linewidth}{!}{%
\begin{tabular}{l c c c c c}
\hline
\textbf{Dataset} & \textbf{Model Size} & \textbf{Sequence Absence} & \textbf{Activity Absence} & \textbf{Sequence Mismatch} & \textbf{Activity Mismatch} \\ \hline
Small            & 7B                 & 0.0007                   & 0.0007                   & 0.3567                    & 0.3633                     \\ 
Small            & 13B                & 0                        & 0                        & 0.526                     & 0.75                       \\ 
Medium           & 7B                 & 0.0023                   & 0.001                    & 0.5254                    & 0.367                      \\ 
Medium           & 13B                & 0.0003                   & 0.0007                   & 0.4693                    & 0.7687                     \\ 
Large            & 7B                 & 0.001                    & 0.0003                   & 0.3141                    & 0.3433                     \\ 
Large            & 13B                & 0.0003                   & 0                        & 0.56                      & 0.7538                     \\ \hline
\end{tabular}%
}
\end{table*}

Comparing the performance of sequence and activity diagrams reveals that activity diagrams generally exhibit fewer syntax errors after fine-tuning. Baseline models perform better on sequence diagrams initially, as their simpler structure reduces the likelihood of errors. However, sequence diagrams are more susceptible to corrupted outputs caused by syntax errors, often producing black or green error images. These corrupted outputs degrade the SSIM scores significantly, even when the BLEU scores remain relatively high. In contrast, fine-tuned models excel at handling the complexity of activity diagrams. The improvements observed in syntax error rates and diagram mismatch scores demonstrate the effectiveness of larger datasets and fine-tuning techniques. The 13B LoRA-tuned model achieves the lowest error rates overall, reinforcing the importance of both model size and training strategy in improving UML diagram generation accuracy.

\subsection{Result Overview}

The experiments confirmed that larger models and datasets improve BLEU and SSIM metrics, reduce syntax errors, and enhance diagram accuracy. Larger models, particularly the 13B configurations, outperformed smaller models across most metrics, demonstrating lower syntax error rates and greater fidelity to ground truth diagrams. Sequence diagrams achieved higher BLEU scores, but their SSIM scores suffered due to syntax errors that degraded visual fidelity. Being more complex, activity diagrams benefited significantly from fine-tuning, exhibiting reduced syntax errors and improved image quality. Fine-tuned models trained on larger datasets consistently demonstrated better generalization and alignment with reference diagrams. Fine-tuning eliminated UML absence errors and reduced diagram mismatches, with only baseline models displaying non-zero UML absence scores.

LoRA fine-tuning yielded results comparable to full fine-tuning but revealed trade-offs related to hardware constraints. While LoRA fine-tuning significantly reduces the number of trainable parameters and lowers memory consumption, our findings highlight an important trade-off: increased training time and higher total FLOPs compared to full fine-tuning. Despite requiring fewer GPUs (two versus four for full fine-tuning), LoRA exhibited slower step rates and longer convergence times. This was largely due to hardware and batch size constraints. The LLaVA training pipeline is designed to maintain a consistent global batch size. To match the original global batch size of 128 (as used in the official 8-GPU setup), we configured LoRA training on two GPUs with a per-device batch size of 16 and four accumulation steps. In contrast, standard fine-tuning used four GPUs with a batch size of 8 per device and the same accumulation steps. While the global batch size was preserved, using fewer GPUs with larger per-device batches increased memory pressure and reduced per-step efficiency. For example, across the 13B model on extra-large activity datasets, LoRA fine-tuning consumed approximately $2.772 \times 10^{15}$ FLOS over 9 hours, compared to $2.48 \times 10^{15}$ FLOS in 5.95 hours for full fine-tuning. This suggests that LoRA's efficiency gains in memory and GPU requirements come at the cost of computational throughput---likely due to differences in batch size and limited GPU resources. However, the flexibility it offers---especially for resource-constrained setups---often outweighs the extended runtime, making it a valuable strategy for scalable domain adaptation.

Although real-world data was not used during training, we evaluated model generalization on a small set of naturally sourced UML diagrams. The results, while showing small improvements with training, revealed a notable performance gap compared to synthetic test data. The discrepancy is likely due to differences in naming structure and semantic coherence, as the synthetic data was generated using randomized labels. Nonetheless, the experiment demonstrates the feasibility of using synthetic data for this task and highlights opportunities for improving generalization. The extensive synthetic datasets reveal that, regardless of other factors, with sufficient data the models can effectively learn to represent UML code from images.

\section{Discussion} 
This study investigates the role of the LLaVA-1.5 model in automating UML diagram-to-code generation, expanding on prior work by Conrardy and Cabot. While their research prioritizes syntactic precision and prompt design for models like GPT-4 and CogVLM \citep{conrardy2024, cogvl2024}, our broader approach asses both structural fidelity and computational efficiency. By incorporating BLEU and SSIM metrics, we evaluate textual and visual accuracy, offering a more comprehensive performance assessment across various datasets and model configurations. Results emphasize the critical role of dataset scale and fine-tuning strategies. Larger LoRA-tuned models, particularly the 13B configurations, consistently outperform baseline models by achieving higher BLEU and SSIM scores. These models excel at generating structurally accurate UML diagrams, effectively reducing syntax errors and mismatches. Fine-tuning, especially with LoRA, demonstrates how targeted parameter updates enhance performance without overwhelming computational costs, making advanced models accessible to resource-limited environments.

Compared to Conrardy and Cabot’s focus on prompt refinement, our study addresses computational scalability, a key factor in adoption. While they achieved impressive results with structural accuracy for class diagrams, we highlight the importance of scalability and model adaptability for complex behavioral diagrams like sequence and activity workflows. Both studies demonstrate the growing potential of multimodal models in software engineering. Still, our inclusion of error analysis—quantifying mismatches and syntax issues—offers deeper insights into areas where models can improve further. Our research also reveals that model misinterpretations, such as defaulting to incorrect diagram types, are more prevalent in smaller datasets. However, fine-tuned models correct these issues by adhering more closely to prompts and minimizing mismatches. These findings demonstrate that larger datasets and well-tuned models enable MM-LLMs to handle dynamic workflows effectively. By addressing key challenges such as structural errors, prompt misinterpretations, and computational bottlenecks, our work sets the stage for scalable and efficient UML diagram automation.
\subsection{Avenues for Further Study}

The BLEU metric has been a foundational tool for evaluating code generation; however, its reliance on syntactic $n$-gram matching may limit its applicability for UML tasks, where visual fidelity and structural relationships are also critical. CodeBLEU extends BLEU by incorporating abstract syntax tree analysis and data flow consistency, aligning evaluations more closely with functional correctness. However, its current design focuses on traditional programming languages such as Python or C \citep{2020codebleu}. Exploring how similar techniques might be adapted for UML tasks could open avenues for aligning evaluations with textual and visual aspects. Furthermore, approaches such as pass@k, which evaluates correctness across multiple outputs, and frameworks such as EvalPlus, which uses mutation-based testing to identify potential model vulnerabilities \citep{NEURIPS2023_43e9d647}, present interesting opportunities for advancing UML evaluation methods by integrating textual and diagrammatic assessments. 

Emerging multimodal models such as LLaVA-NeXT \citep{liu2024llavanext} and LLaVA-OneVision \citep{li2024llavaonevision}, which incorporate advanced OCR, multi-resolution processing, and cross-modal knowledge transfer, offer promising directions for improving UML automation by better handling various types of diagram and maintaining accuracy at varying resolutions. While our experiments centered on LLaVA-based models, we acknowledge the importance of incorporating additional benchmark models to broaden the comparative landscape. Our primary objective was to demonstrate the feasibility of using a multimodal LLM for large-scale UML code generation. Within this scope, LLaVA-1.5 at both 7B and 13B scales proved sufficient for validating our approach. Nevertheless, including models such as Deepseek R1 \citep{deepseekai2025}, Claude, or CogVLM in future work would offer a more comprehensive perspective on architectural differences, visual reasoning capabilities, and generalization. 

Our models achieved strong performance on synthetic datasets, gains on more realistic UML samples were notably more modest. As illustrated in Figs. \ref{fig:realworld_activity_performance} and \ref{fig:realworld_sequence_performance}, fine-tuned models demonstrated only marginal improvements over the baseline when evaluated on real-world diagrams. This discrepancy stems from synthetic diagrams' structure, which fail to fully capture the artifacts commonly found in typical UML diagrams. To bridge the realism gap, future iterations of our synthetic dataset could categorize diagrams into common application domains such as banking, e-commerce, authentication workflows, and web development. The thematic structuring would align synthetic diagrams more closely with real-world use cases, thereby improving their semantic fidelity. Additionally, incorporating domain-specific terminology and realistic naming conventions within these categories would further enhance the authenticity of the synthetic data. Nevertheless, our results strongly suggest that synthetic data can serve as a powerful training signal with sufficient volume and variation: MM-LLMs exhibit a clear capacity to learn to accurately reconstruct UML code from images. The result highlight a promising direction---by further enhancing the visual realism and variability of synthetic training data, future models may bridge the domain gap and achieve high-fidelity UML code generation even in real-world scenarios.

\section{Conclusion}
This study introduces a scalable and efficient framework to automate UML diagram-to-code generation, directly addressing a critical need in software engineering. We demonstrate how fine-tuned multimodal large language models adapt to dynamic workflows and event-driven processes by focusing on sequence and activity diagrams. Integrating LoRA fine-tuning improves accuracy while reducing computational demands, making the approach accessible for resource-constrained environments. Our dual-metric evaluation framework combines BLEU for textual accuracy and SSIM for visual fidelity, creating a robust benchmark for assessing UML generation models. This approach ensures that generated outputs maintain structural precision and syntactic correctness, bridging the gap between diagrammatic design and machine-readable code. Fine-tuned models consistently outperform baseline versions by achieving higher fidelity and reducing structural misalignment errors.

The study highlights opportunities for future exploration, such as modifying advanced metrics like CodeBLEU and pass@k, which offer deeper insights into functional correctness, and leveraging state-of-the-art models like LLaVA-NeXT and LLaVA-OneVision to enhance reasoning and scalability. Expanding datasets to include additional UML types, such as class diagrams, and incorporating synthetic and real-world examples will strengthen model generalization and robustness. These advancements will refine UML automation tools that align with the evolving requirements of modern software engineering workflows. Our contributions demonstrate the potential of intelligent, scalable AI systems to automate complex design tasks and bridge theoretical advances with practical applications. The framework offers software engineering solutions and broader fields requiring structured and dynamic representations, including healthcare, education, and industrial automation. Establishing a strong foundation for AI-driven design and implementation creates new opportunities for innovation across diverse industries, advancing intelligent system integration and automation.

\section*{Acknowledgments}

The authors thank the University of Oklahoma's Supercomputing Center for Education \& Research for providing access to their supercomputing resources. We also thank the Data Institute for Societal Challenges for facilitating GPU resources through OSCER's infrastructure. In addition, we acknowledge AFWERX for their support and contributions. Finally, the authors extend their deepest gratitude to Dr. Andrew H. Fagg for his invaluable feedback on the thesis that formed the foundation of this work. His guidance were helped shape the direction and quality of the research. 

\section*{CRediT authorship contribution statement}

\textbf{Averi Bates:} Methodology, Software, Validation, Formal analysis, Investigation, Data curation, Writing - original draft, Writing - review \& editing, Visualization. \textbf{Ryan Vavricka:} Data curation, Software, Writing - review \& editing. \textbf{Shane Carleton:} Conceptualization, Supervision, Project administration, Funding acquisition. \textbf{Ruosi Shao:} Writing - review \& editing. \textbf{Chongle Pan:} Conceptualization, Supervision, Project administration, Writing - review \& editing.

\section*{Funding Statement}

This project was funded by the U.S. National Science Foundation under the Accelerating Research Translation program 2331409 and the U.S. Air Force Small Business Innovation Research program. Additional funding for publication was provided by the University of Oklahoma Libraries' Open Access Fund

\section*{Declaration of Competing Interest}

Shane Carleton reports financial support was provided by US Department of the Air Force. Ryan Vavricka reports financial support was provided by MapLarge. Shane Carleton reports a relationship with MapLarge that includes: employment. Averi Bates has patent issued to University of Oklahoma. Shane Carleton has patent issued to University of Oklahoma. Chongle Pan has patent issued to University of Oklahoma. If there are other authors, they declare that they have no known competing financial interests or personal relationships that could have appeared to influence the work reported in this paper.

\section*{Data Availability}

The training and testing UML data supporting the findings of this study are available and can be found at \href{https://doi.org/10.5281/zenodo.15103682}{https://doi.org/10.5281/zenodo.15103682}. Please note that this excludes any associated prompts. Access to the data will be granted for non-commercial research purposes.

\appendix

\section{Supplementary Figures}
The following is the supplementary material related to this article.
\renewcommand{\thefigure}{S\arabic{figure}}
\setcounter{figure}{0}

\begin{figure}[htbp]
    \centering
    \includegraphics[width=0.6\textwidth]{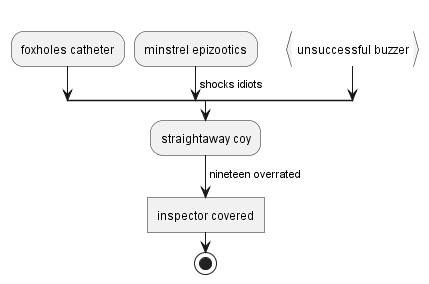}
    \caption{Activity diagram showing decision-making and workflows.}
    \label{fig:activity_code_combo}
\end{figure}

\begin{figure}[htbp]
    \centering
    \includegraphics[width=0.48\textwidth]{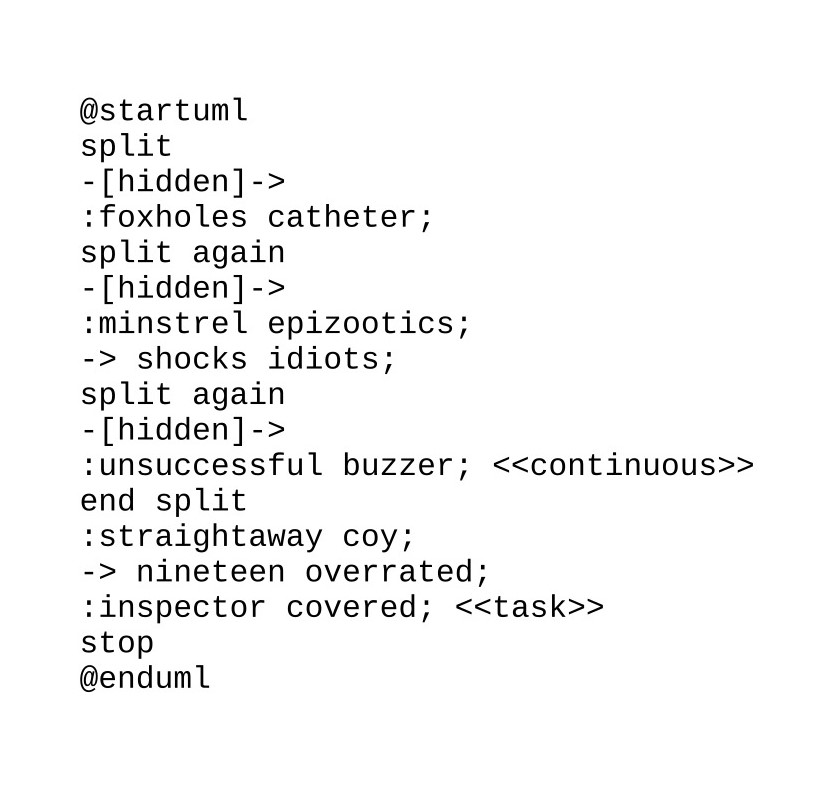}
    \caption{PlantUML code generating the activity diagram.}
    \label{fig:activity_code}
\end{figure}

\begin{figure}[htbp]
    \centering
    \begin{small}
    \begin{lstlisting}
    {
        "id": "unique-id-1234",
        "image": "path/to/image.jpg",
        "conversations": [
            {"from": "human", "value": "<image>\n Generate the most 
            likely UML code from the diagram."},
            {"from": "gpt", "value": "Expected UML code output here..."}
        ]
    }
    \end{lstlisting}
    \end{small}
    \caption{Example JSON data format used for model training.}
    \label{fig:json_example}
\end{figure}

\begin{figure}[htbp]
    \centering
    \includegraphics[width=0.35\textwidth]{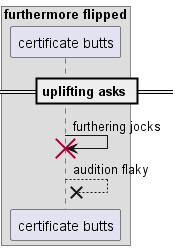}
    \caption{Ground truth UML diagram used for comparison.}
    \label{fig:high_score_true}
\end{figure}

\begin{figure}[htbp]
    \centering
    \includegraphics[width=0.4\textwidth]{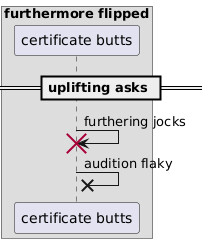}
    \caption{Generated UML sequence diagram with high SSIM and BLEU scores, showing strong structural and textual alignment with the ground truth.}
    \label{fig:high_score_generated}
\end{figure}

\begin{figure}[htbp]
    \centering
    \includegraphics[width=0.6\textwidth]{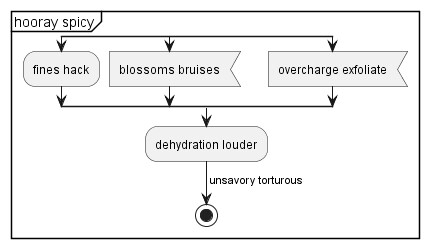}
    \caption{Ground truth UML diagram illustrating the original reference structure.}
    \label{fig:low_score_true}
\end{figure}

\begin{figure}[htbp]
    \centering
    \includegraphics[width=0.4\textwidth]{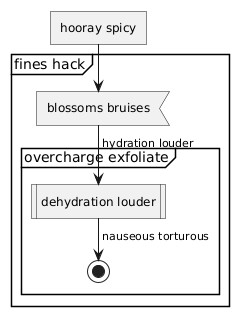}
    \caption{Generated UML activity diagram with low SSIM and BLEU scores, highlighting structural misalignments and textual discrepancies compared to the ground truth.}
    \label{fig:low_score_generated}
\end{figure}

\begin{figure}[htbp]
    \centering
    \includegraphics[width=0.55\textwidth]{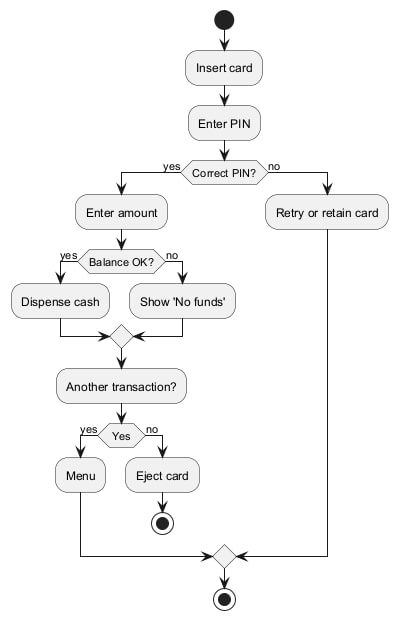}
    \caption{Ground truth UML diagram illustrating the original reference structure of real world data rather than synthetic.}
    \label{fig:low_score_true_realworld_act}
\end{figure}

\begin{figure}[htbp]
    \centering
    \includegraphics[width=0.55\textwidth]{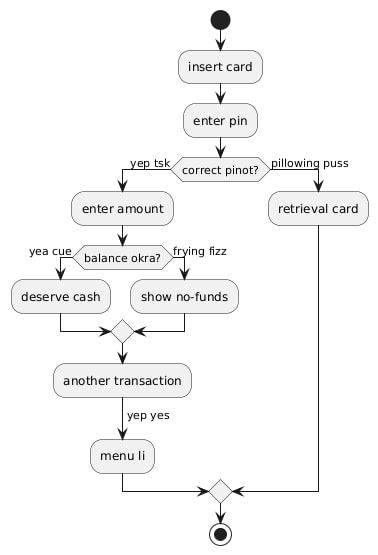}
   \caption{Example of a generated UML activity diagram with a moderatly high SSIM score and low BLEU score, highlighting structural and syntactic issues that arise from training on unstructured synthetic data.}
    \label{fig:low_score_generated_realworld_act}
\end{figure}

\begin{figure}[htbp]
    \centering
    \includegraphics[width=0.333\textwidth]{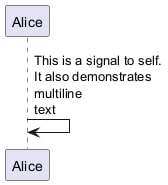}
    \caption{Ground truth UML diagram illustrating the original reference structure of real world data rather than synthetic.}
    \label{fig:low_score_true_realworld_act}
\end{figure}

\begin{figure}[htbp]
    \centering
    \includegraphics[width=0.333\textwidth]{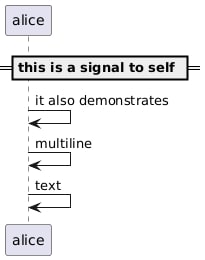}
   \caption{Example of a generated UML sequence diagram with low SSIM and BLEU scores, highlighting structural and syntactic issues that arise from training on unstructured synthetic data.}
    \label{fig:low_score_generated_realworld_act}
\end{figure}
\clearpage

\bibliographystyle{elsarticle-harv} 
\bibliography{example}

\end{document}